\documentclass[prb,twocolumn,aps,superscriptaddress,floatfix]{revtex4}
\usepackage{amsmath}
\usepackage{amssymb}
\usepackage{amsfonts}
\usepackage{bm}
\usepackage{graphicx}
\usepackage{subfigure}
\usepackage{color}
\usepackage[svgnames]{xcolor}
\usepackage{soul}
\usepackage{hyperref}
\usepackage{verbatim}% Long comments
\usepackage{float}

\begin{document}

\title{Moir{\'e}-like superlattice generated van Hove singularities in strained CuO$_2$ double layer}

\author{A.O.~Sboychakov}
\affiliation{Institute for Theoretical and Applied Electrodynamics, Russian Academy of Sciences, 125412 Moscow, Russia}

\author{K.I.~Kugel}
\affiliation{Institute for Theoretical and Applied Electrodynamics, Russian Academy of Sciences, 125412 Moscow, Russia}
\affiliation{Institute of Metal Physics, Ural Branch, Russian Academy of Sciences, 620990 Ekaterinburg, Russia}

\author{A.~Bianconi}
\affiliation{Rome International Center for Materials Science Superstripes RICMASS, via dei Sabelli 119A, 00185 Rome, Italy}
\affiliation{Institute of Crystallography, CNR, Via Salaria Km 29.300, 00015 Monterotondo Rome, Italy}
\affiliation{National Research Nuclear University, MEPhI (Moscow Engineering Physics Institute), Kashirskoe sh. 31, 115409 Moscow, Russia}

\begin{abstract}
While it is known that the double-layer Bi$_2$Sr$_2$CaCu$_2$O$_{8+y}$ (BSCCO) cuprate superconductor exhibits a one-dimensional (1D) incommensurate superlattice (IS), the effect of IS on the electronic structure remains elusive. Following the recent shift of the interest from underdoped to optimum and overdoped phase in BSCCO by increasing the hole doping $x$, controlled by the oxygen interstitials concentration $y$, here we focus on the multiple splitting of the density of states (DOS) peaks and emergence of higher order van Hove singularities (VHS) due to the 1D incommensurate superlattice. It is known that 1D incommensurate wave vector $\textbf{q} = \epsilon \textbf{b}$ (where $\textbf{b}$ is the reciprocal lattice vector of the orthorhombic lattice) is controlled by the misfit strain between different atomic layers in the range $0.209$\,--\,$0.215$  in BSCCO and in the range $0.209$\,--\,$0.25$ in Bi$_2$Sr$_2$Ca$_{1-x}$Y$_x$Cu$_2$O$_{8+y}$ (BSCYCO). This work reports the theoretical calculation of a complex pattern of VHS due to the 1D incommensurate superlattice with large 1D quasi-commensurate supercells with the wave vector $\epsilon =9/\eta$ in the range $36 > \eta > 43$. The similarity of the complex VHS splitting and appearing of higher order VHS in a mismatched CuO$_2$ bilayer with VHS due to the moir{\'e} lattice in strained twisted bilayer graphene is discussed. This makes mismatched CuO$_2$ bilayer quite promising for constructing quantum devices with tuned physical characteristics.
\end{abstract}

%\pacs{73.20.At}

%73.22.Pr  Electronic structure of graphene
%73.20.At	 Surface states, band structure, electron density of states

\date{\today}

\maketitle

\section{Introduction}

Different van der Waals heterostructures~\cite{LiuNatRevMat2016}, such as bilayer graphene~\cite{RozhkovPhRep2016} and transition metal dichalcogenide bilayers~\cite{WestonNatNTech2020,ZhaoNatMat2021} provide an actively progressing testing ground for a variety of quantum phenomena, e.g., flat bands, electron nematicity, etc. Indeed, if we put together two layers of van der Waals materials, a lattice mismatch or twist angle lead to the formation of moir\'e patterns with large supercells. This provides a promising means of the band and crystal structure engineering leading, in particular, to the moir\'e modulated topological order~\cite{TongNatPh2017}, and paving way to such novel field of research as twistronics~\cite{RenChinPh2020}.

Bilayers of atomic cuprate oxide superconductors appear to be not less interesting. The lattice parameter and aperiodic local lattice distortions associated with variable microstrain are not key physical parameters either in BCS and unconventional superconductivity theories. Nevertheless, a compelling evidence for the key role of CuO$_2$ lattice in the mechanism of high-temperature superconductivity and charge density wave (CDW) phenomena in cuprate perovskites has been reported~\cite{KrockenbergerACSOmega2021,ParkProgSCCr2021,ConradsonPNAS2020,
SederholmCondMat2021,JangPRX2017}. These findings confirm early experimental results provided by local and fast experimental methods based on the use of synchrotron X-ray radiation ~\cite{BianconiPRL1996,SainiEPJB2003,BianconiPRB1996,BianconiPRB1996a,
DiCastroEPJB2000} pointing toward the relevant role of the micro-strain in superconducting cuprate perovskites~\cite{BianconiJPCM2000,AgrestiniJPhA2003}.

The Bi$_2$Sr$_2$CaCu$_2$O$_{8+y}$ (BSCCO) crystal~\cite{BianconiPRB1996,BianconiPRB1996a,DiCastroEPJB2000} is a misfit layer compound~\cite{WiegersProgSSCh1996} exhibiting an incommensurate composite structure~\cite{WiegersProgSSCh1996,IntTabl2004} and large atomic displacements from average positions in the [BiO], [SrO], [CuO$_2$], and [Ca] layers. These displacements form an incommensurate composite structure, where a texture of nanoscale insulating and metallic atomic stripes, see for example~\cite{EtrillardPRB2000,GrebillePRB2001comment,BianconiSSC1994,BeskrovnyyJPhConfSer2012} Bi$_2$Sr$_2$Ca$_{1-x}$Y$_x$Cu$_2$O$_{8+y}$ (BSCYCO) with six different elements in the average crystalline unit cell forming an archetypal case of High Entropy Perovskites (HEPs). The search of new HEPs is today a hot topic focusing on the manipulation of the complexity, where the entropy control plays a key role in the determination of material functionalities like it is observed in bismuth 2212 iron ferrite and ferrate compounds~\cite{PerezPRB1997,SedykhPhC2003}, thermoelectric misfit layered cobaltite oxide perovskites~\cite{LambertJSSCh2001,GrebilleAcCrB2007} also in the form of 2D flakes~\cite{KimJAP2012} taking advantage of the polymorphism of perovskite structure with competing nanoscale crystallographic phases~\cite{GavrichkovJPhChL2019,MazzaJVacSci2022}.

Spatially correlated incommensurate lattice stripy modulations in Bi$_2$Sr$_2$CaCu$_2$O$_{8+y}$ superconductor can now be visualized by the scanning submicron X-ray diffraction focusing on the superlattice satellites of the main X-ray reflections~\cite{KanekoAPL2004,PocciaPRB2011,PocciaPRM2020}. The results show that, while oxygen concentration and the doping level can change depending on the sample treatments and thermal history, the incommensurate superlattice wave vector is controlled by the misfit strain and it exhibits a remarkable stability. The critical temperature exhibits a dome as a function of oxygen concentration or by potassium deposition~\cite{ZhangSciBul2016}, but the maximum critical temperature of the dome ~\cite{BianconiPocciaJSCNM2012} is controlled by the lattice stripe aperiodic structure~\cite{BianconiJpnJAP1993} and misfit strain~\cite{BianconiJPCM2000}.

Recently the scientific interest turned toward the normal state of cuprate perovskites exhibiting high-temperature superconductivity in the overdoped regime in the proximity of the superconductor-to-metal transition tuned by doping, while for many years the focus was on the underdoped regime. The investigations of the overdoped phase~\cite{AlldredgeNatPh2008,PasupathyScience2008,GomesNature2007,
BozovicNature2016,HePRX2021,DrozdovNature2018,LiNJP2018,MaierPRRes2020} of Bi2212 have found evidence that the superconducting gap appearing at temperatures well above the critical temperature $T_c$, while below $T_c$  a large fraction of the normal state remains uncondensed. Scanning tunneling microscopy has shown that on the edge of the overdoped regime, the gap remains in isolated puddles above $T_c$, so that superconducting puddles are intercalated by normal metal phase. Moreover, at the same time, the Fermi surface topology changes giving rise to an electronic topological Lifshitz transition. This provides compelling evidence for a nanoscale phase separation giving a granular landscape, where superconducting puddles are embedded in a normal phase. The phase separation scenario was predicted to appear at the Lifshitz transition several years ago~\cite{KaganPhRep2021,RakhmanovJETPL2017,KugelPRB2008,BianconiSST2015,KugelSST2008}.

It is known that the interlayer interaction between CuO$_2$ of BSCCO planes produces a bilayer splitting of the 2D Cu(3$d$)O(2$p$) single band into two bands with even and odd symmetry exhibiting a double-peaked logarithm divergent density of states (DOS) due to standard van Hove singularities (VHS) in a 2D metal. However, it is not known how the 1D superstructure controls the VHS.

While many electronic band structure calculations of the average structure BSCCO have been published in these last 33 years (see e.g., the papers beginning from Ref.~\onlinecite{MassiddaPhC1988} up to one of the most recent papers, Ref.~\onlinecite{NokelainenPRB2020}), to our knowledge, the electronic band structure calculations of BSCCO with the incommensurate superstructure with the wave vector $0.209 < q <0.215$ (see e.g., Ref.~\onlinecite{HewatPhC1989}) known since 1989, has never been reported. Here, we report the band structure calculation of the strained CuO$_2$ bilayer with the incommensurate structure focusing on the high doping range, where correlations become negligible and the van Hove singularity crosses the chemical potential. We focus on the variation of the electronic topological transition from a hole Fermi surface to an electron Fermi surface in the overdoped regime in the BSCCO double layer perovskite taking into account the aperiodic incommensurate lattice modulation.

In Bi$_2$Sr$_2$CaCu$_2$O$_{8+y}$, the mobility of oxygen interstitials in the CuO$_2$ planes is high in the temperature range 200--380 K, and it is  possible to control the concentration $y$ of oxygen interstitials to change the hole doping $x$ in the layered perovskite.

The insulating Bi$_2$Sr$_2$YCu$_2$O$_8$ exhibits a commensurate modulation with $\varepsilon=0.25$, which indicates the formation of lattice stripes with commensurate period $n =1/\varepsilon = 4$ lattice units. The superconducting Bi$_2$Sr$_2$CaCu$_2$O$_{8+y}$ exhibits a commensurate modulation with $\varepsilon=0.209$, which indicate the formation of lattice stripes superlattice with incommensurate period $n =1/0.209=9/43$ lattice units. For $\varepsilon=0.2$, the CuO$_2$ lattice should be decorated by a commensurate stripes superlattice with $n =1/\varepsilon=1/0.2=5$ lattice units. For intermediate values of the superlattice wave vector $0.25 < \varepsilon < 0.2$, the incommensurate superstructure~\cite{BakRPPh1982} consists of the mixture of the $\varepsilon = 1/n$ and $\varepsilon = 1/m$ order, i.e. alternating the so called $n=4$ stripes portions and $m=5$ stripes portions. Indeed, quasi-commensurate periods $\lambda = 1/\varepsilon = (nx+my)/(n+m)$   intermediate between two main commensurate wave vectors $1/n$ and $1/m$ with periods of $m$ and $n$ lattice units, respectively, occur at integer numbers of lattice unit cells  $\eta=(n+m)/\varepsilon$. Each quasi-commensurate phase (QCP) corresponds to a modulation wave locked with the underlying lattice onto a rational number. The sequence of quasi-commensurate phases is called the devil's staircase~\cite{BakRPPh1982}, which have been observed in many complex materials.

The quasi-commensurate modulation period in BSCCO is given by $1/\varepsilon=(4x+5y)/(4+5)$, where $x+y=9$. Therefore, $\eta = 9/\varepsilon=4x+5y$ varies in the range $36< \eta <45$ and at each integer value of $\eta$ the system reaches a quasi-commensurate phase at each step of a {\it devil's staircase}. We have found that in the BSCCO samples, where the chemical potential is tuned by the concentration of oxygen interstitials, the mismatched CuO$_2$ bilayer is tuned by the misfit strain at the quasi-commensurate modulation with $\eta=43$ i.e. $\epsilon$=$9/43= 0.209$ in agreement with the recent scanning micro-X-ray diffraction \cite{PocciaPRM2020}, which is the {\it devil's staircase} for
 $x = 2$  and $y =5$. Therefore, we have $4\times2+5\times7= 43$ (i.e. a quasicrystal made of 8 insulating lattice units and 35 metallic lattice units).

\begin{figure*}[t]
\centering
\includegraphics[width=0.3\textwidth]{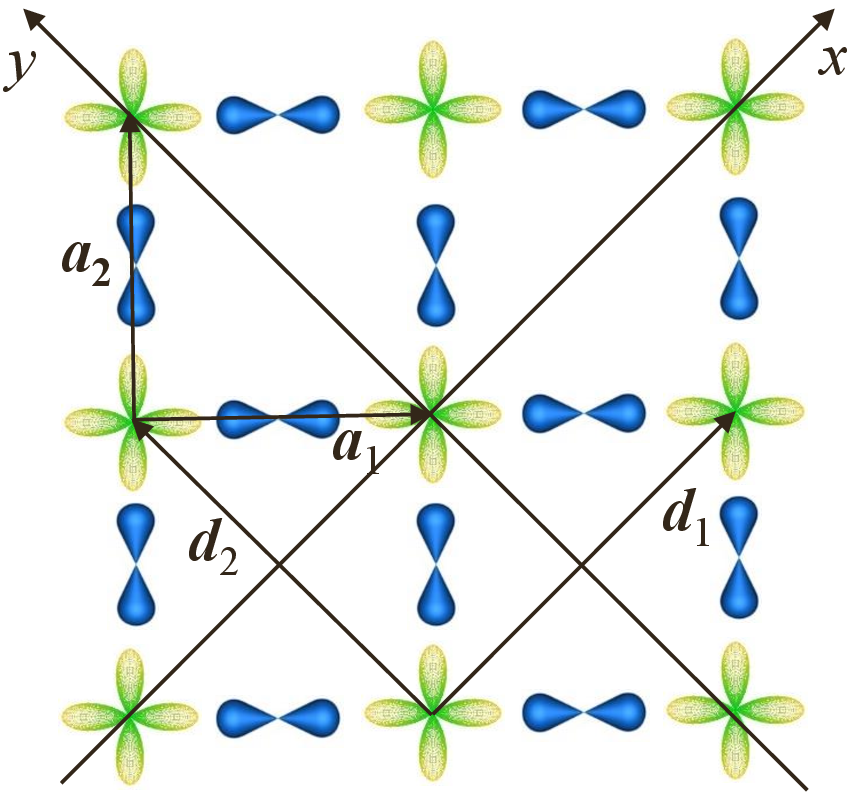}\hspace{5mm}%\vspace{5mm}\\
\includegraphics[width=0.3\textwidth]{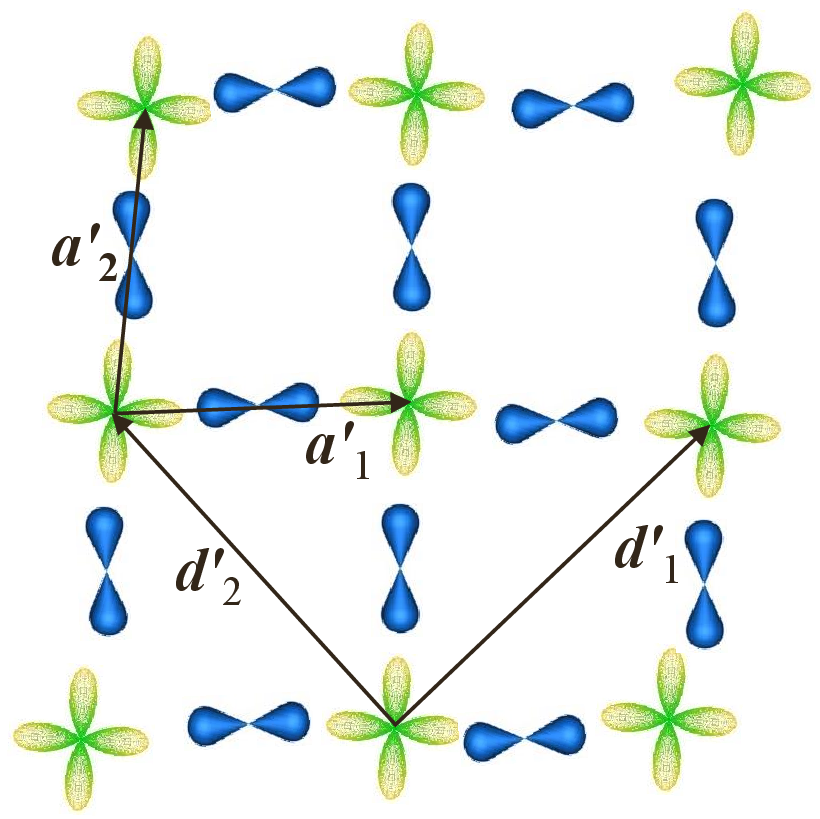}
\includegraphics[width=0.36\textwidth]{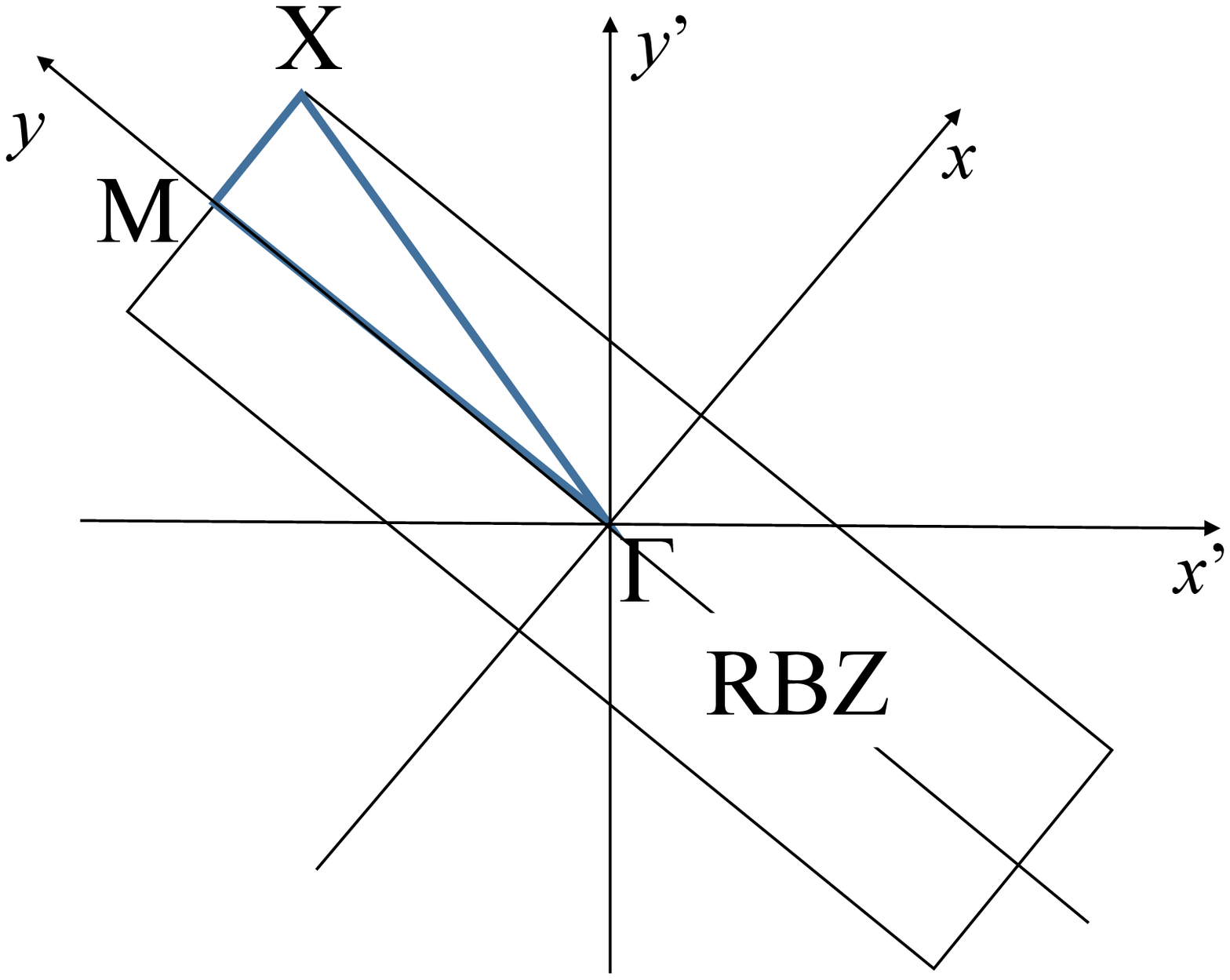}
\caption{Schematics of the CuO$_2$ layer 1 (left panel) and 2 (middle panel). Layer 1 is unstretched. Layer 2 is either stretched or compressed in diagonal ($x$) direction. As a result, for some commensurate stretching/compression, a superstructure appears. The superlattice contains $2N_1$ elementary unit cells of the layer 1 and $2N_2$ elementary unit cells of the layer 2. Right figure shows the sketch of the reduced Brillouin zone of the superlattice. The $\Gamma$ point is in the coordinate origin, the $M$ point has coordinates $\mathbf{M}=(0,\,\pi/d)$, while coordinate of the $X$ point is $\mathbf{X}=(\pi/(N_1d),\,\pi/d)$. The thick blue triangle shows the contour, along which the spectra (see below) are calculated. \label{FigLayers}}
\end{figure*}

\section{Geometry of the system} \label{geom}

We model the system under study as consisting of two mismatched CuO$_2$ layers. The layer 1 is assumed to be quadratic. For the reason described below, we choose the reference frame as shown on the left panel in Fig.~\ref{FigLayers}. For such a choice, the vectors connecting next nearest-neighbor Cu ions have coordinates $\mathbf{d}_1=d(1,\,0)$ and $\mathbf{d}_2=d(0,\,1)$, where $d$ is the distance between nearest copper ions in the {\it diagonal} direction. We choose~\cite{BianconiPRB1996} $d=5.37$\,\AA.  We also introduce vectors connecting nearest-neighbor Cu ions of the layer 1 (lattice vectors of the unit cell of layer 1). They are expressed via vectors $\mathbf{d}_{1,2}$ according to $\mathbf{a}_1=(\mathbf{d}_1-\mathbf{d}_2)/2$ and $\mathbf{a}_2=(\mathbf{d}_1+\mathbf{d}_2)/2$. The positions of Cu ions in the layer 1 are
\begin{equation}
\mathbf{r}^{d}_{1\mathbf{n}}\equiv\mathbf{r}_{1\mathbf{n}}=n\mathbf{a}_1+m\mathbf{a}_2\,,
\end{equation}
where $\mathbf{n}=(n,\,m)$ ($n$, $m$ are integers), while the positions of oxygen ions are
\begin{equation}
\mathbf{r}^{p_x}_{1\mathbf{n}}=\mathbf{r}_{1\mathbf{n}}+\frac12\mathbf{a}_1\,,
\;\;\mathbf{r}^{p_y}_{1\mathbf{n}}=\mathbf{r}_{1\mathbf{n}}+\frac12\mathbf{a}_2\,.
\end{equation}

Layer 2 is assumed to be stretched or compressed in the $x$ (diagonal) direction. As a result, the vectors connecting next nearest-neighbor Cu ions in this layer have coordinates $\mathbf{d}'_1=d(1-\delta,\,0)$ and $\mathbf{d}'_2=d(0,\,1)$, where parameter $\delta$ describes the strength of the stretching ($\delta<0$) or compression ($\delta>0$). The vectors connecting nearest-neighbor Cu ions of the layer 2 become $\mathbf{a}'_1=(\mathbf{d}'_1-\mathbf{d}'_2)/2$ and $\mathbf{a}'_2=(\mathbf{d}'_1+\mathbf{d}'_2)/2$. The positions of copper and oxygen ions in the layer 2 are
\begin{eqnarray}
\mathbf{r}^{d}_{2\mathbf{n}}&\equiv&\mathbf{r}_{2\mathbf{n}}=
n\mathbf{a}'_1+m\mathbf{a}'_2\,,\;\nonumber\\
\mathbf{r}^{p_x}_{2\mathbf{n}}&=&\mathbf{r}_{2\mathbf{n}}+\frac12\mathbf{a}'_1\,,\;
\mathbf{r}^{p_y}_{2\mathbf{n}}=\mathbf{r}_{2\mathbf{n}}+\frac12\mathbf{a}'_2\,.
\end{eqnarray}

For a rational value of $1-\delta = N_1/N_2$ ($N_1$ and $N_2$ are co-prime positive integers), the system has a superstructure. To satisfy the periodicity conditions, we must include two adjacent chains of copper and oxygen ions into superlattice cell aligned in the diagonal ($x$) direction. As a result, the superlattice cell will contain $2N_1$ copper ions of layer 1 and $2N_2$ copper ions of layer 2. It also will contain $4N_1$ oxygen ions of layer 1 and $4N_2$ oxygen ions of layer 2. The superlattice vectors are $\mathbf{R}_1=N_1\mathbf{d}_1=N_2(1-\delta)\mathbf{d}_1$ and $\mathbf{R}_2=2\mathbf{d}_2$. The superlattice Brillouin zone (reduced Brillouin zone, RBZ) has a shape of the rectangle with the size $2\pi/(N_1d)$ in the $x$ direction and $2\pi/d$ the $y$ direction (see the right panel of Fig.~\ref{FigLayers}).

\section{Tight-binding model of mismatched CuO$_2$ bilayer} \label{tight_binding}

\begin{figure}[t]
\centering
\includegraphics[width=0.99\columnwidth]{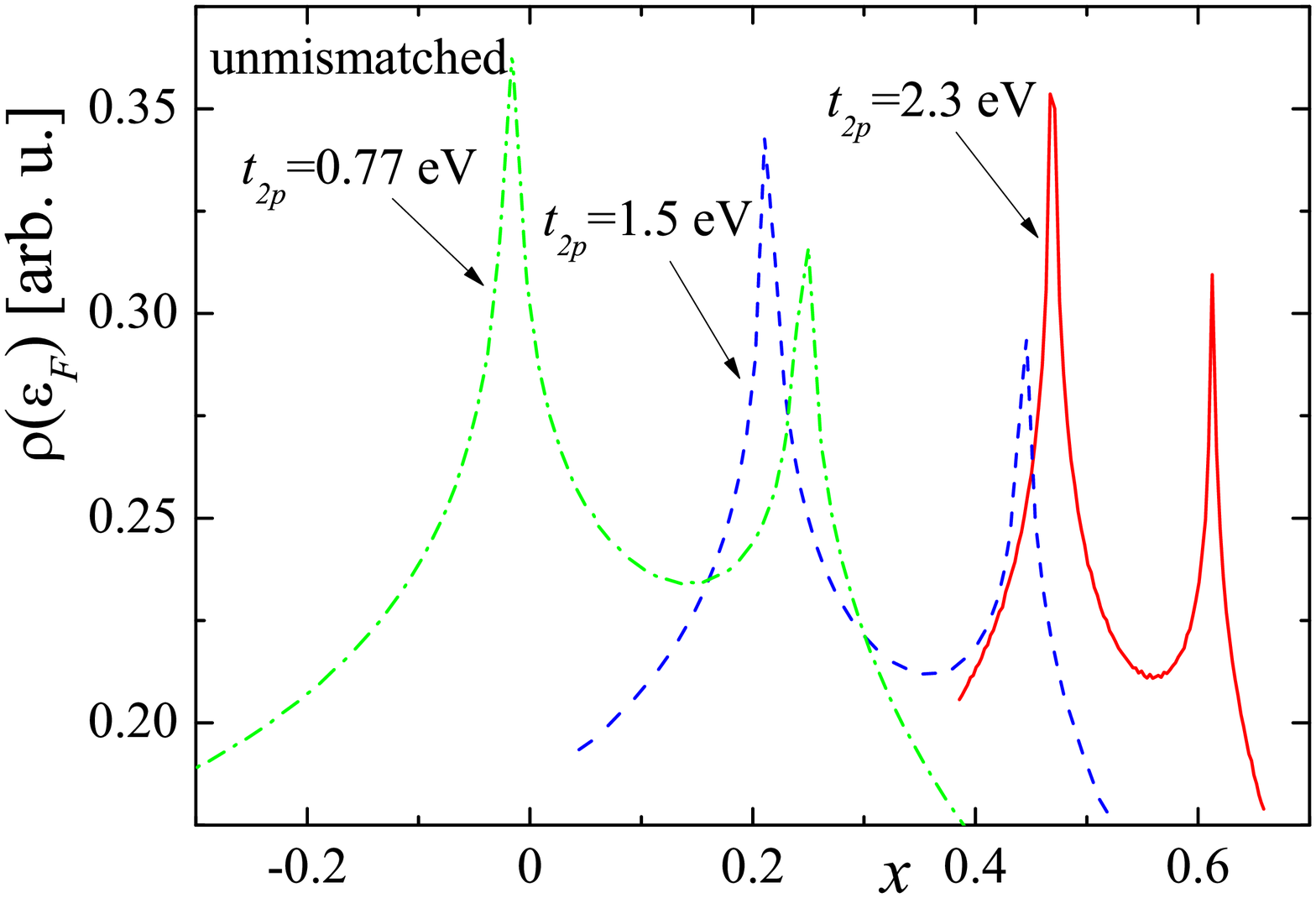}
\caption{Densities of states at the Fermi level for unmismatched layers as functions of doping $x$, calculated for three different values of $t_{2p}$. Other model parameters are: $\varepsilon_{x^2-y^2}=0$, $\varepsilon_{p_x}=\varepsilon_{p_y}=-0.9$\,eV, $\varepsilon_a=6.5$\,eV, $t_{1p}=1.6$\,eV, $t_0=0.27$\,eV, and $t^{aa}_0=0.75$\,eV. In order to avoid logarithmic singularities, the densities of states are averaged over the energy range $\Delta E=7$\,meV.}\label{FigDOStsp}
\end{figure}

To describe electronic properties of mismatched CuO$_2$  bilayer, we use the four-band tight-binding model. The model includes two types of $p$ oxygen orbitals: $p_x$ orbitals of oxygen ions located in positions $\mathbf{r}^{p_x}_{i\mathbf{n}}$ and $p_y$ orbitals of oxygen ions located in positions $\mathbf{r}^{p_y}_{i\mathbf{n}}$ . Model also includes two types of $d$ orbitals of copper located in positions $\mathbf{r}^{d}_{i\mathbf{n}}$:  these are $x^2-y^2$ and `axial' ($a$) orbitals. We do not specify here the nature of the latter $a$ orbital, and only mention its circular symmetry in $xy$ plane. It can be, e.g., $s$ or  $3z^2-r^2$ orbital of copper, or their superposition. We include this orbital  into the model because it gives the largest effect into interlayer hybridization~\cite{LDAenergyBands1995}.

\begin{figure*}[t]
 \centering
\includegraphics[width=0.32\textwidth]{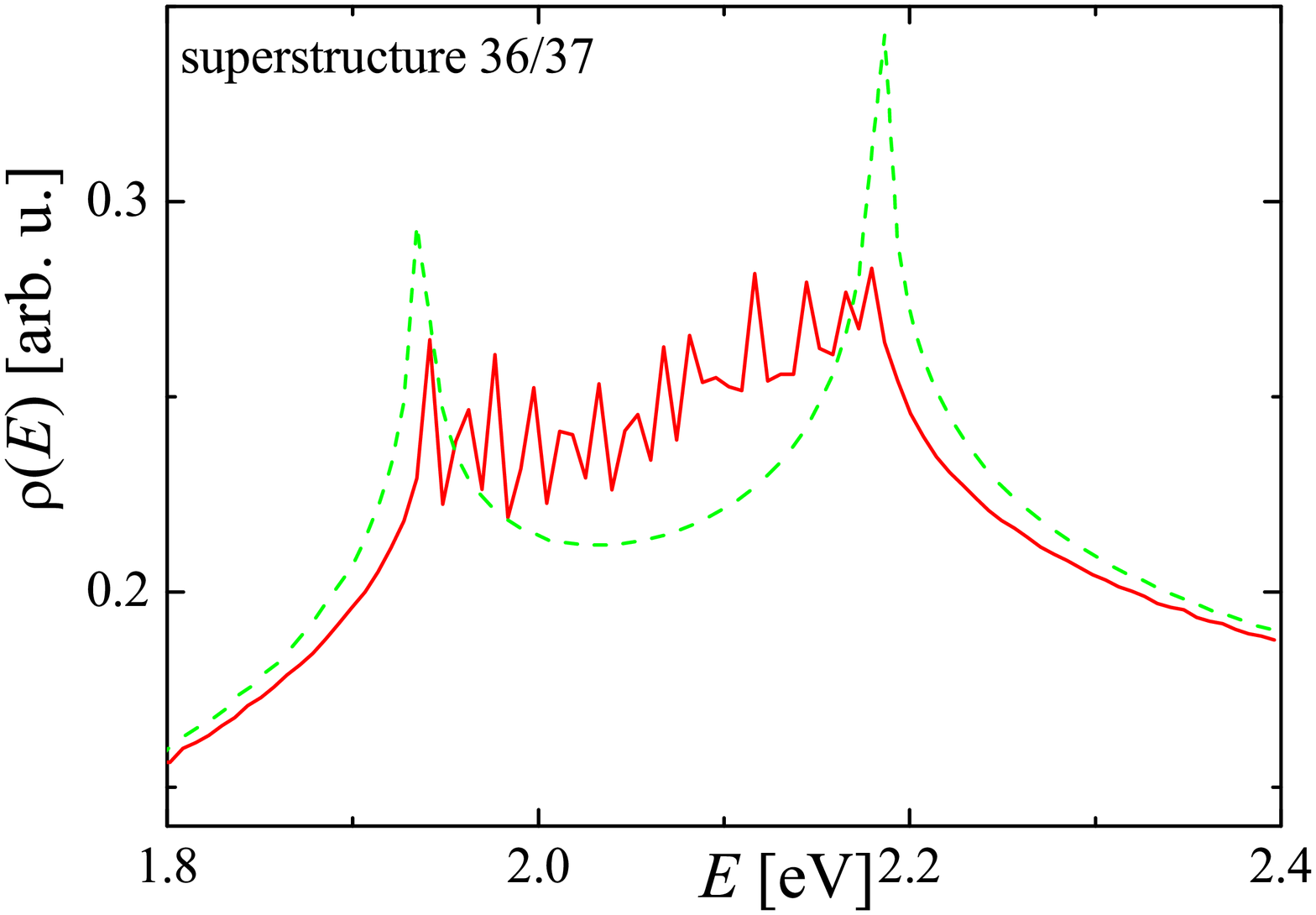}
\includegraphics[width=0.32\textwidth]{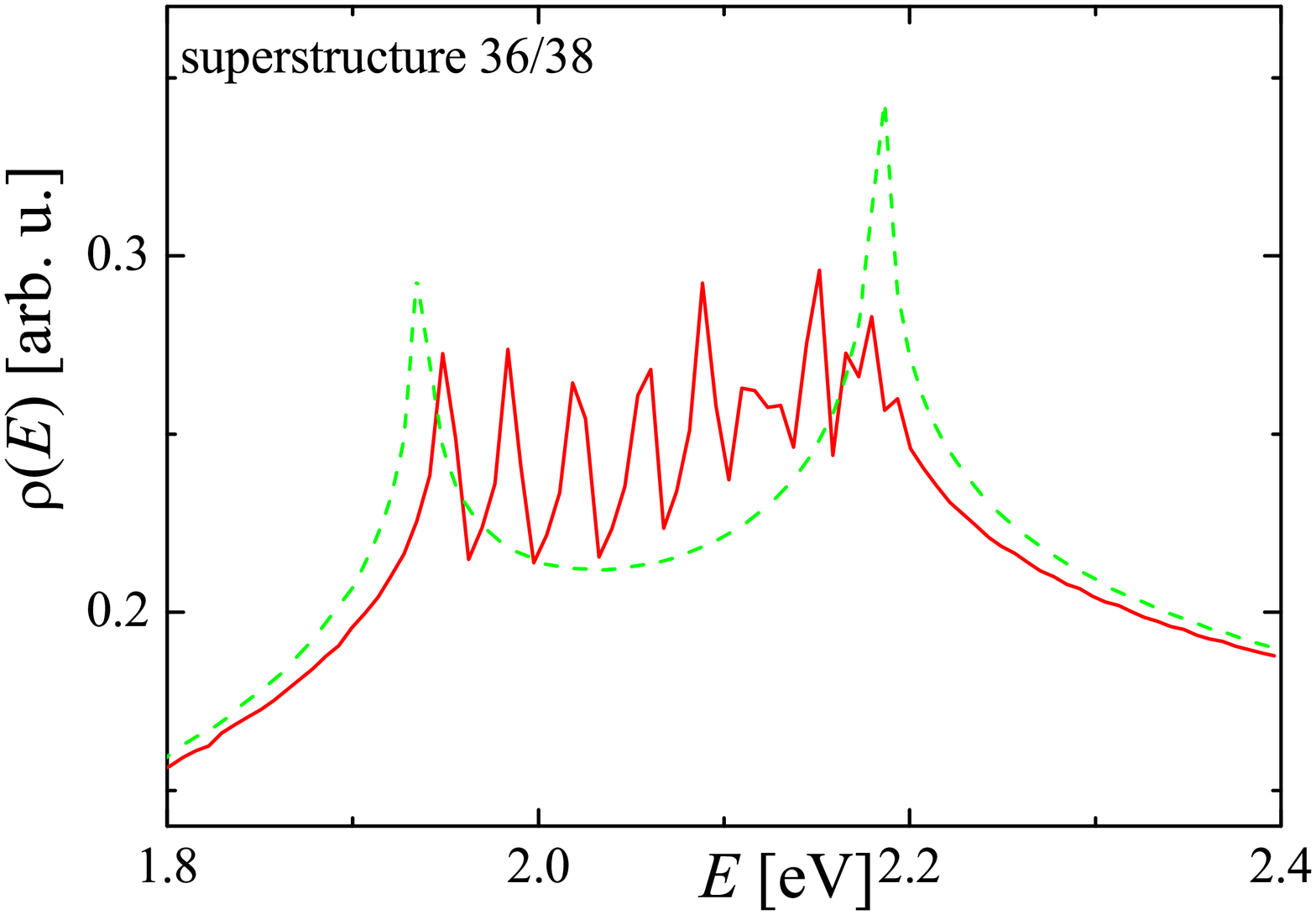}
\includegraphics[width=0.32\textwidth]{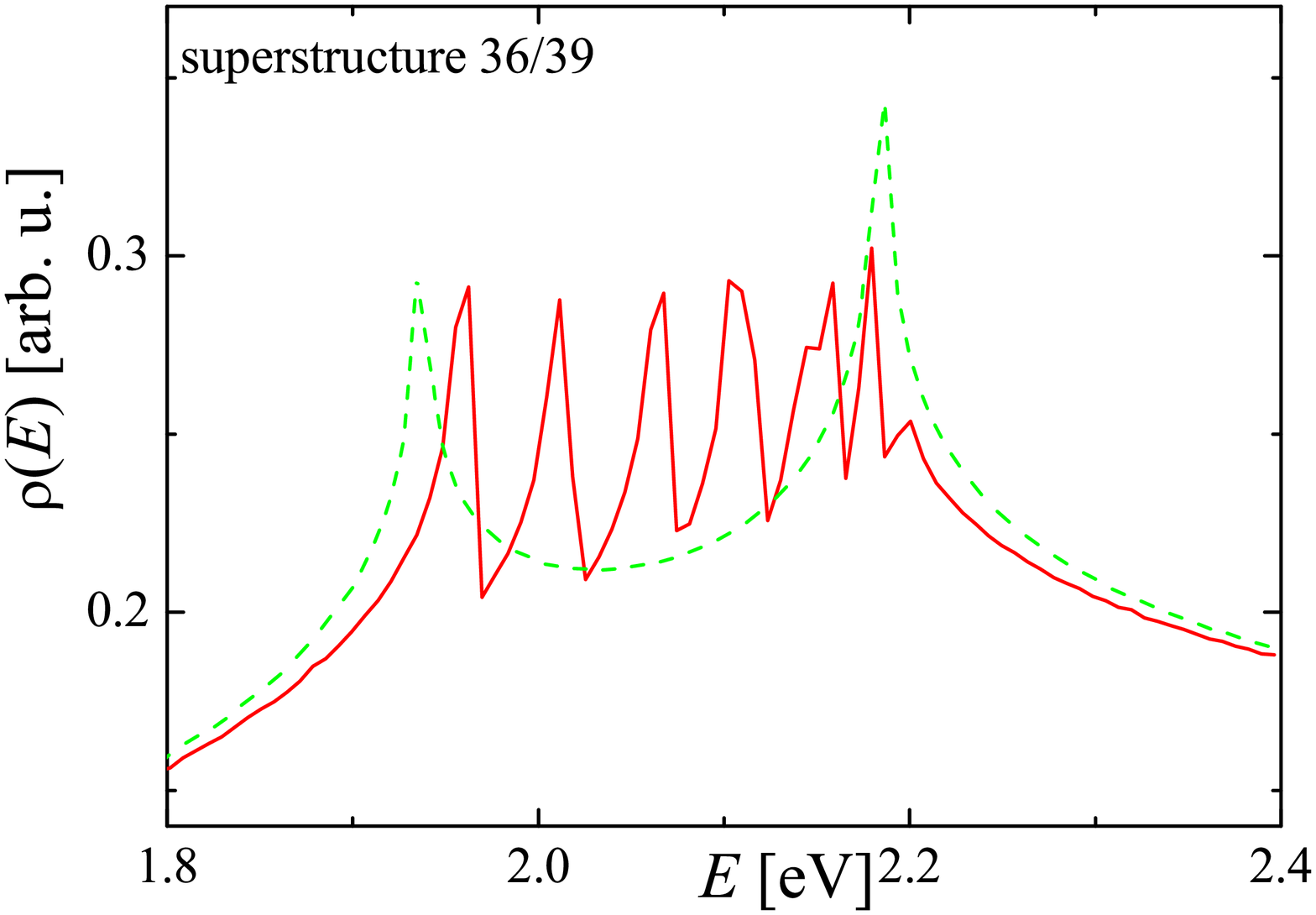}\vspace{5mm}
\includegraphics[width=0.32\textwidth]{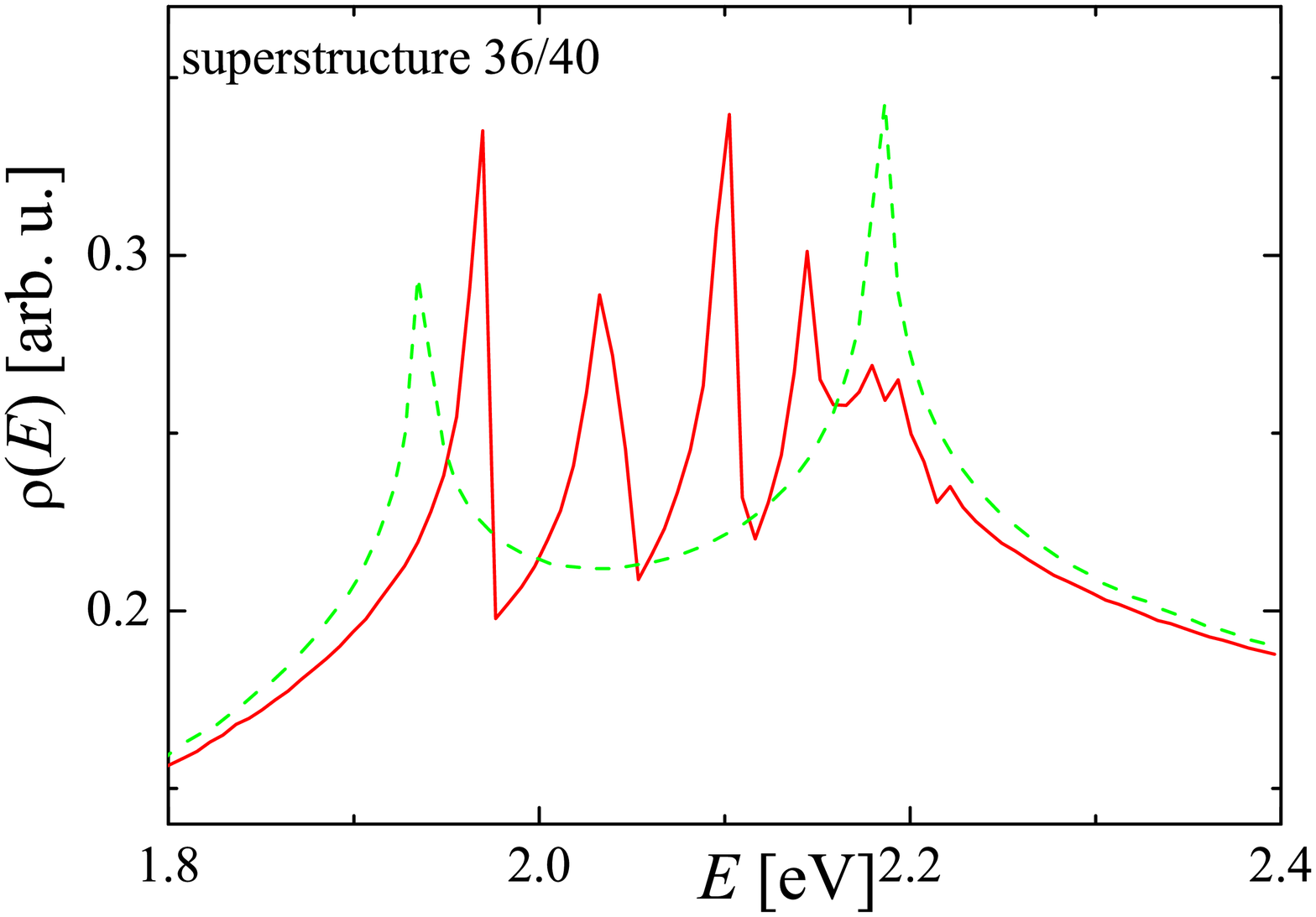}
\includegraphics[width=0.32\textwidth]{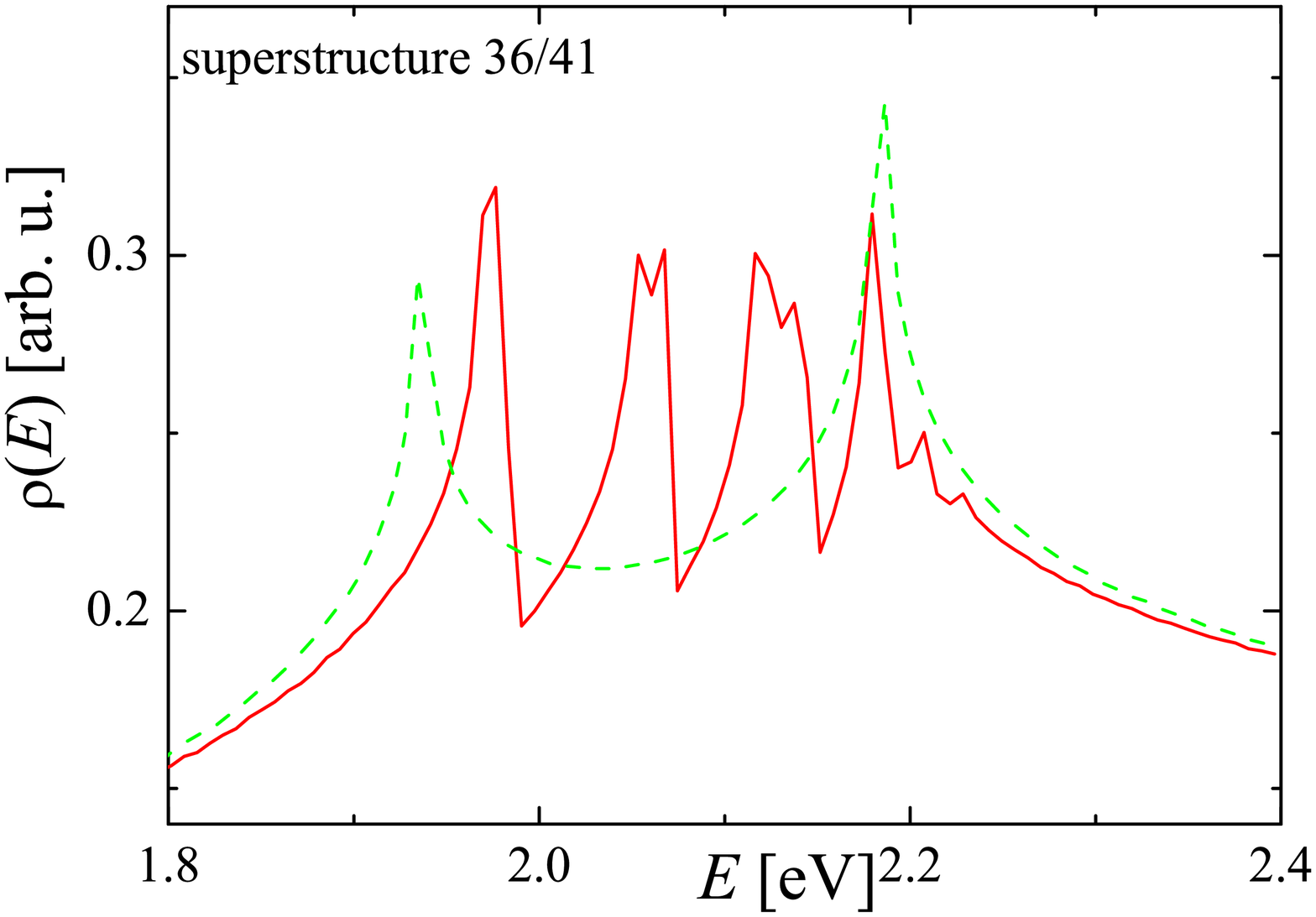}
\includegraphics[width=0.32\textwidth]{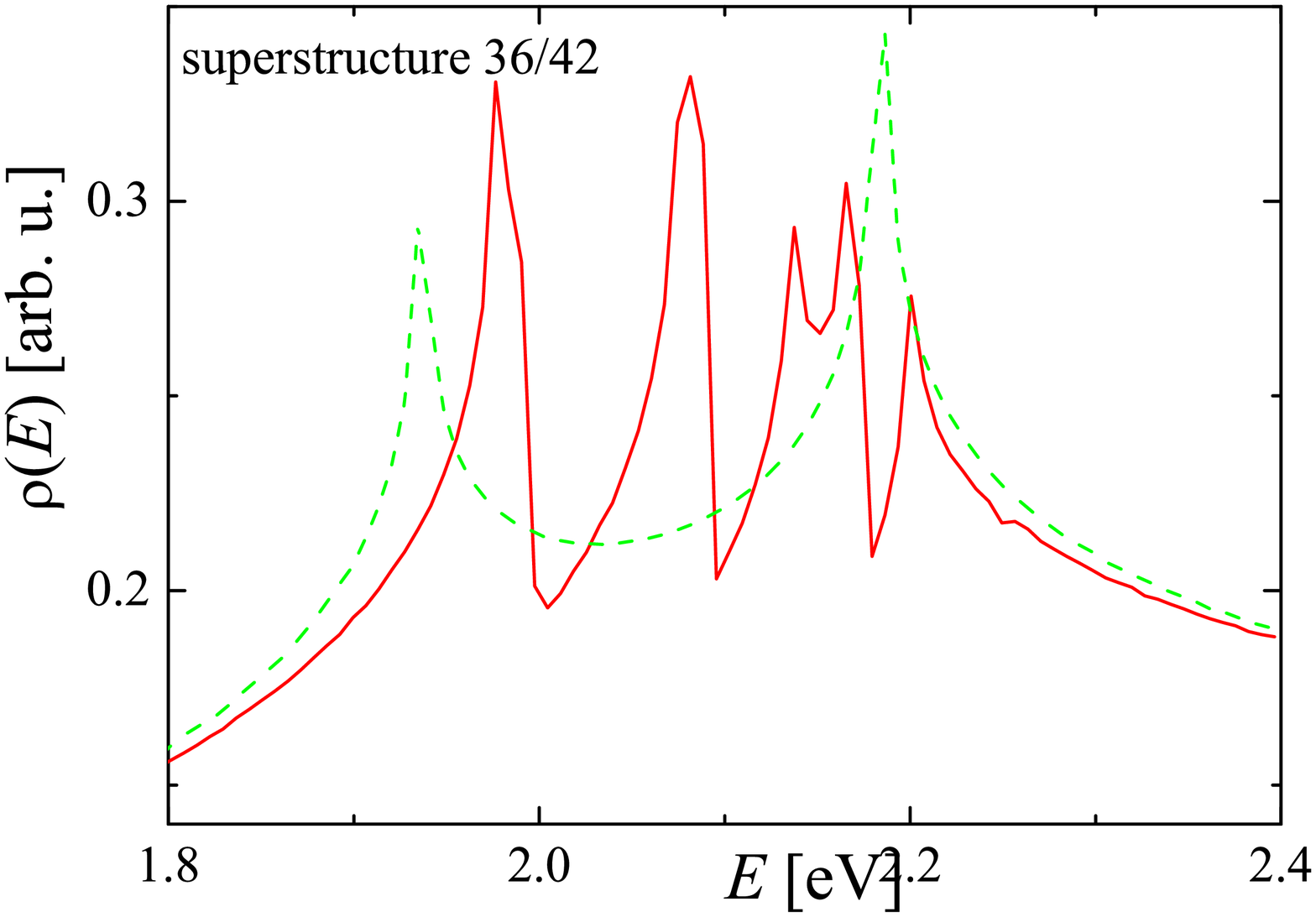}\vspace{5mm}
\includegraphics[width=0.32\textwidth]{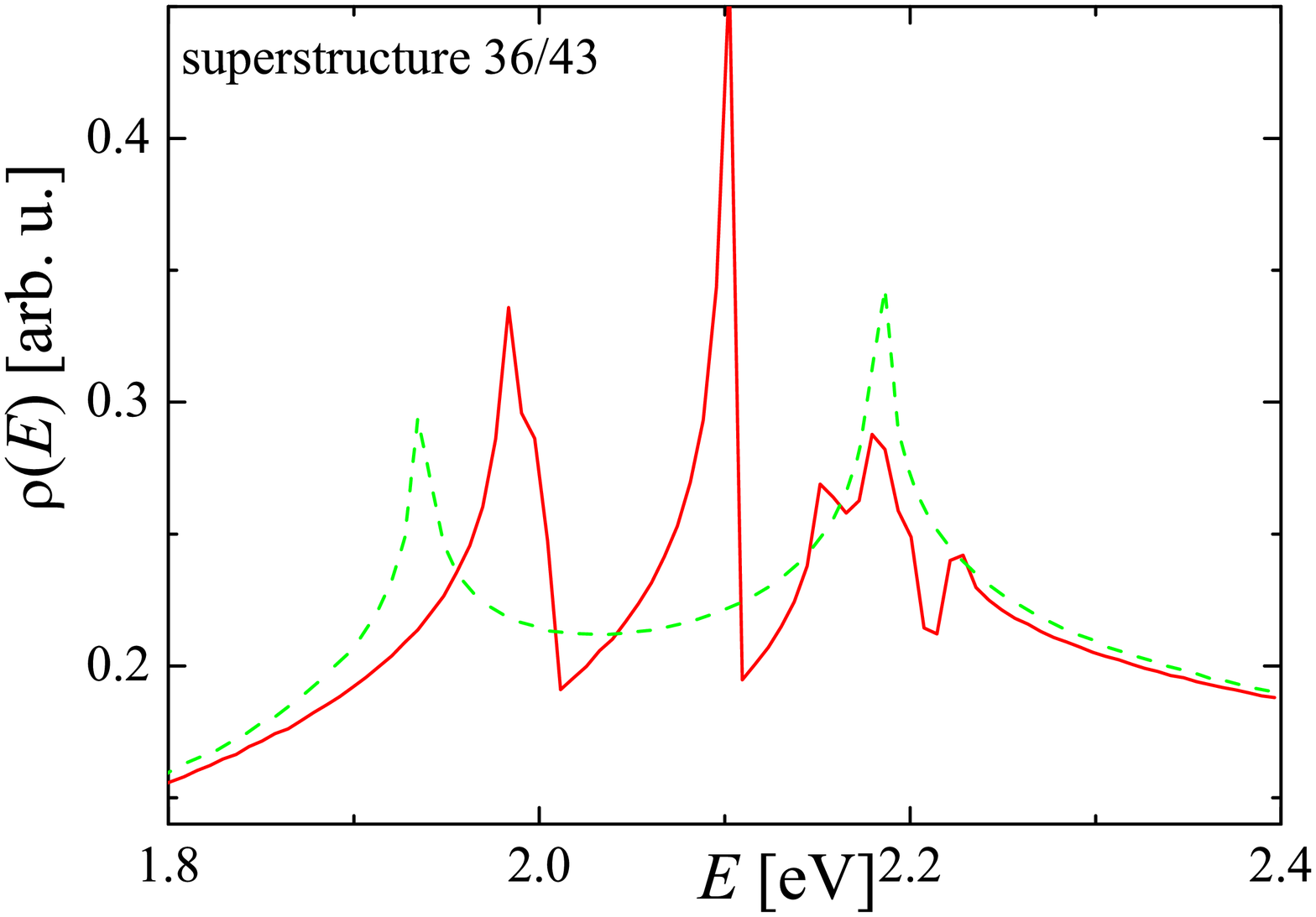}
\includegraphics[width=0.32\textwidth]{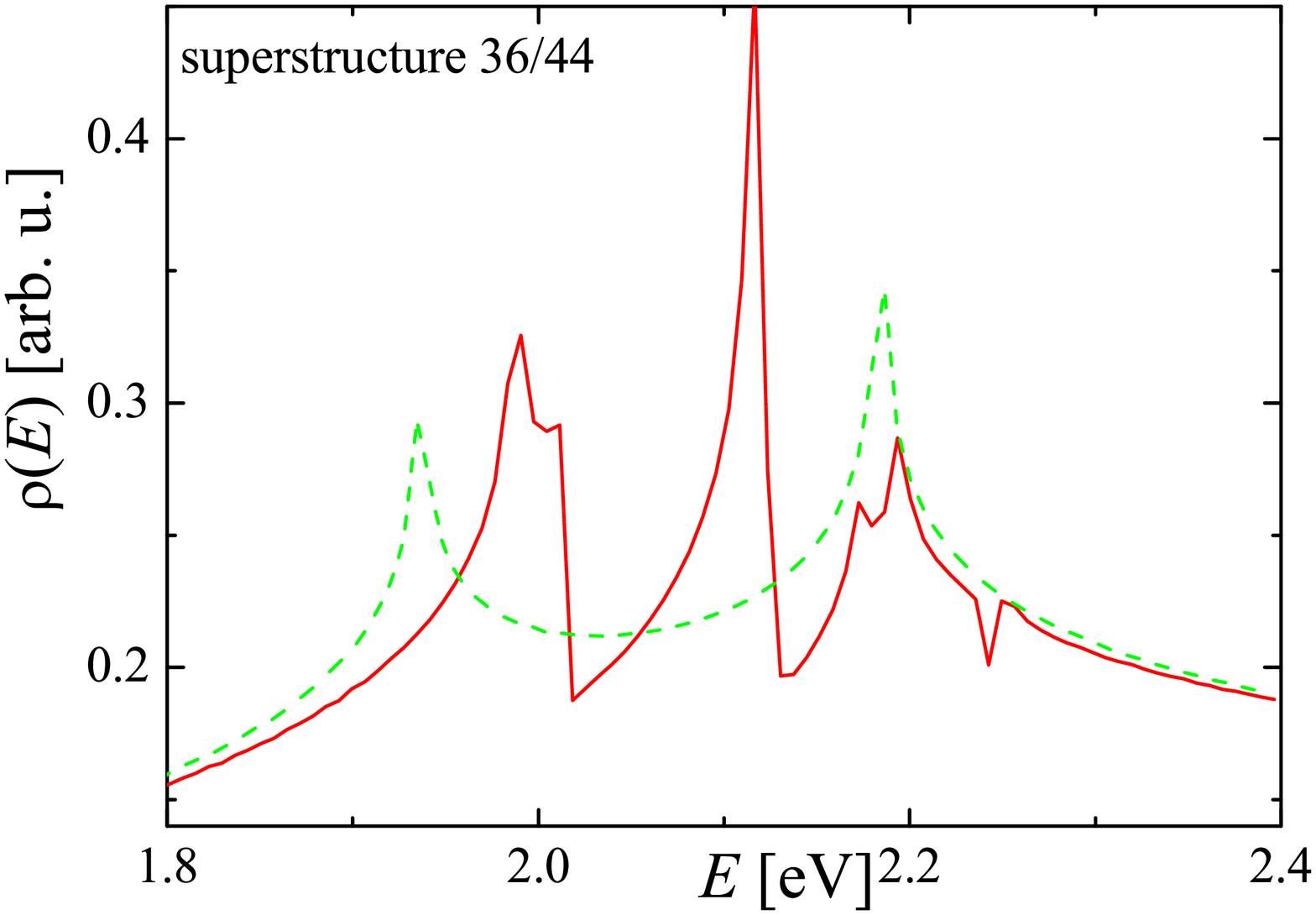}
\includegraphics[width=0.32\textwidth]{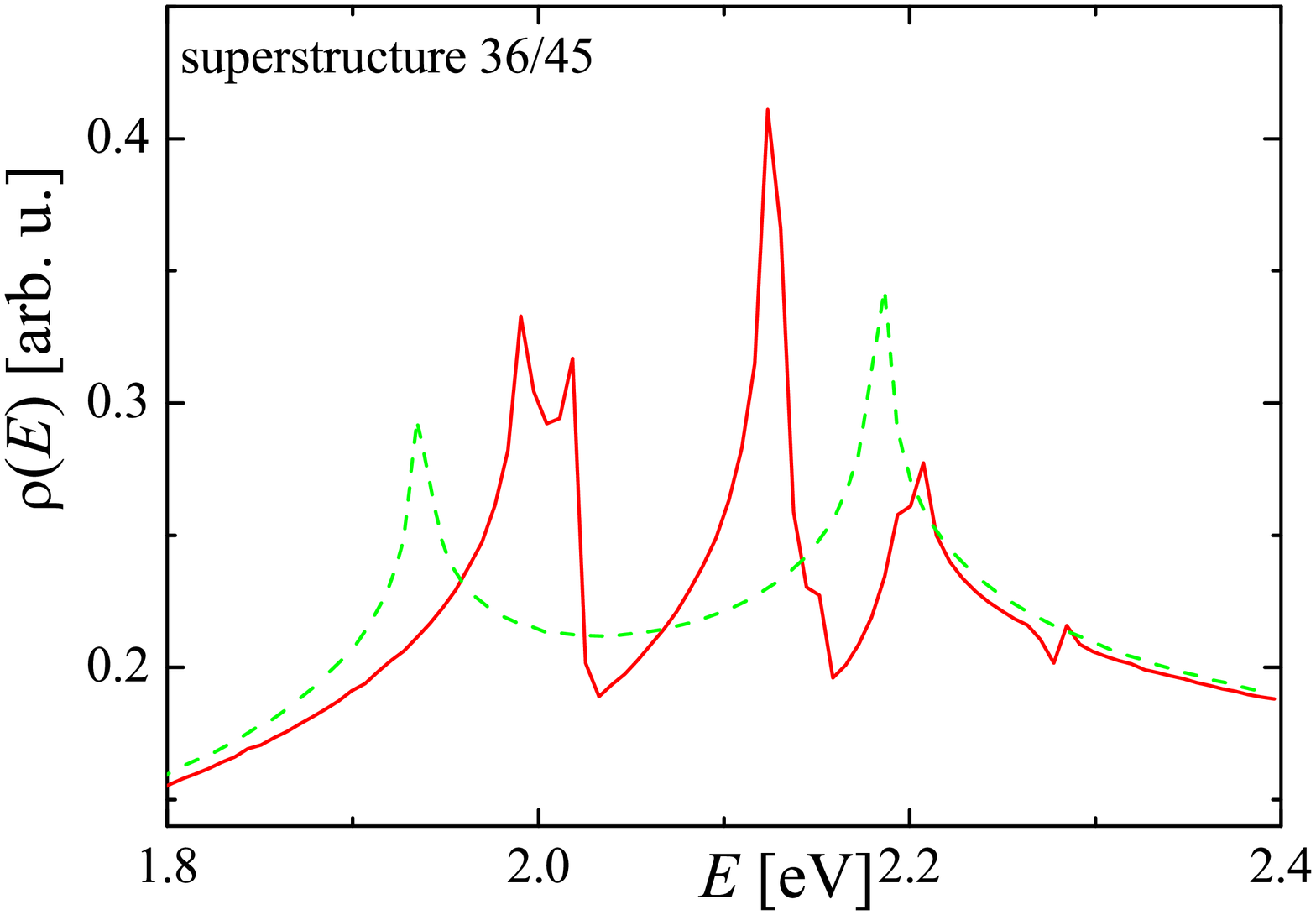}
\caption{Densities of states as functions of energy calculated for nine superstructures with $N_1=36$ and $N_2$ varying from $37$ to $45$. Model parameters correspond to Fig.~\ref{FigDOStsp} with $t_{2p} = 1.5$ eV. In order to avoid logarithmic singularities, the densities of states are averaged over the energy range $\Delta E=7$\,meV. The dashed green curve in each plot corresponds to unmismatched layers.}\label{FigDOS36_X}
\end{figure*}

The Hamiltonian under study can be written as
\begin{equation}
H=H_1+H_2+H_{12}\,.
\end{equation}
The terms $H_{1,2}$ correspond to each individual layer, while $H_{12}$ describes the interlayer hybridization. Let us consider first intralayer terms. We use the simplest intralayer tight-binding Hamiltonian including only nearest neighbor hoppings of electrons between nearest copper and oxygen ions. The Hamiltonian $H_i$ ($i=1,\,2$) reads
\begin{equation}\label{H1}
H_i=\sum_{\mathbf{n}A\sigma}\varepsilon_{A}d^{\dag}_{\mathbf{n}iA\sigma}
d^{\phantom{\dag}}_{\mathbf{n}iA\sigma}+
\!\sum_{\langle\mathbf{nm}\rangle\atop dp\sigma}\!\left(t^{dp}_{i\mathbf{nm}}d^{\dag}_{\mathbf{n}id\sigma}
d^{\phantom{\dag}}_{\mathbf{m}ip\sigma}+H.c.\right),
\end{equation}
where the subscript $A=\{d,\,p\}$ (with $d=x^2-y^2,\,a$ and $p=p_x,\,p_y$) denotes  the considered orbitals, and $d^{\dag}_{\mathbf{n}iA\sigma}$ and $d^{\phantom{\dag}}_{\mathbf{n}iA\sigma}$ are the creation and annihilation operators of electron located in unit cell $\mathbf{n}$ of layer $i$ having orbital index $A$ and spin projection $\sigma$. The first term in~\eqref{H1} includes the local energies of electrons in different orbitals, while the last term describes the nearest-neighbor hopping. The hopping amplitudes $t^{dp}_{i\mathbf{nm}}$ are position dependent. For a given direction of hopping $\mathbf{m}\to\mathbf{n}$, the hopping amplitudes $t^{dp}_{i\mathbf{nm}}$ can we written in the form of $2\times2$ matrix. The relationships between elements of this matrix can be easily obtained from the symmetry analysis of orbitals' wave functions~\cite{SlaterKoster}. Let us introduce the Fourier transformed electron operators $d^{\phantom{\dag}}_{\mathbf{k}iA\sigma}$ and Fourier transformed hopping amplitudes $t^{dp}_{i\mathbf{k}}$. Taking into account the symmetry of the orbitals under study~\cite{SlaterKoster}, one can write for layer 1
\begin{equation}\label{tdp}
t^{dp}_{1\mathbf{k}}=\left(\begin{array}{cc}
-t_{1p}e^{-i\mathbf{ka}_1/2}&t_{1p}e^{-i\mathbf{ka}_2/2}\\
-t_{2p}e^{-i\mathbf{ka}_1/2}&-t_{2p}e^{-i\mathbf{ka}_2/2}\\
\end{array}\right),
\end{equation}
where we introduce two independent Slater--Koster parameters describing the hopping between $x^2-y^2$ and $p$ orbitals ($t_{1p}$) and between $a$ and $p$ orbitals ($t_{2p}$). The similar formula for the layer 2 can be obtained from Eq.~\eqref{tdp} with the change $\mathbf{a}_{1,2}\to\mathbf{a}'_{1,2}$. Strictly speaking, the values of parameters $t_{1p}$ and $t_{2p}$ for layer 2 are different from those for layer 1. In further analysis, however, we neglect this difference due to the smallness of the lattice distortion. Following Ref.~\onlinecite{LDAenergyBands1995}, we use the following values of the tight-binding parameters: $\varepsilon_{x^2-y^2}=0$, $\varepsilon_{p_x}=\varepsilon_{p_y}=-0.9$\,eV, $t_{1p}=1.6$\,eV, $t_{2p}=1.5$\,eV. In our simulations, we use different values of the parameter $\varepsilon_{a}$ describing the on-site energy of $a$ orbital, taking both negative (almost $3z^2-r^2$ orbital) and positive (almost $s$ orbital) values. We also performed simulations for different values of the parameter $t_{2p}$ describing hybridization between $a$ orbitals of copper and $p$ oxygen orbitals. Independent on the values of $\varepsilon_{a}$ and $t_{2p}$, the density of states of individual layer has only one van Hove peak corresponding to the change of the Fermi surface topology. The position of this peak depends on $\varepsilon_{a}$ and $t_{2p}$. In present study, we always focus on the energy range close to the van Hove peak. The obtained results are similar for different values of $\varepsilon_{a}$ and $t_{2p}$.

Let us consider now the interlayer hopping. According to Ref.~\onlinecite{LDAenergyBands1995}, the largest interlayer hopping amplitudes are between $a$ orbitals in two different layers, and between $a$ and $p_{x,y}$ orbitals of different layers. As a result, Hamiltonian $H_{12}$ takes the form
\begin{eqnarray}
H_{12}&=&\!\!\sum_{\mathbf{nm}p\sigma}\!\!\left(t^{ap}_{\bot\mathbf{nm}}
d^{\dag}_{\mathbf{n}2a\sigma}d^{\phantom{\dag}}_{\mathbf{m}1p\sigma}+
t^{pa}_{\bot\mathbf{nm}}d^{\dag}_{\mathbf{n}2p\sigma}
d^{\phantom{\dag}}_{\mathbf{m}1a\sigma}\right)+\nonumber\\
&&\!\sum_{\mathbf{nm}\sigma}\!t^{aa}_{\bot\mathbf{nm}}
d^{\dag}_{\mathbf{n}2a\sigma}d^{\phantom{\dag}}_{\mathbf{m}1a\sigma}+H.c.
\end{eqnarray}

The value of hopping amplitude depends on the mutual positions of two orbitals. In our simulations, we tried different parametrizations of interlayer hopping amplitudes depending on the possible symmetry of $a$ orbital and obtain qualitatively similar results. If $a$ orbital is assumed to be rather of $s$ symmetry, the amplitude $t^{ap_{x,y}}_{\bot\mathbf{nm}}$ of electron hopping from oxygen ion in layer $1$ located in the position $\mathbf{r}^{p_{x,y}}_{1\mathbf{m}}$ to copper ion in layer $2$ located in the position $\mathbf{r}^{d}_{2\mathbf{n}}$ is described by the following Slater--Koster formula
\begin{eqnarray}
t^{ap_{x,y}}_{\bot\mathbf{nm}}=\frac{(\mathbf{r}^{d}_{2\mathbf{n}}
-\mathbf{r}^{p_{x,y}}_{1\mathbf{m}})\mathbf{e}_{x,y}}
{\sqrt{c^2+(\mathbf{r}^{d}_{2\mathbf{n}}-
\mathbf{r}^{p_{x,y}}_{1\mathbf{m}})^2}}V_{\sigma}(\mathbf{r}^{d}_{2\mathbf{n}}
-\mathbf{r}^{p_{x,y}}_{1\mathbf{m}})\,,
\end{eqnarray}
where $\mathbf{e}_{x,y}$ is the unit vector in the direction of $\mathbf{a}_{1,2}$ and the function
\begin{equation}
V_{\sigma}(r)=t_0\sqrt{1+8c^2/d^2}\,e^{-\frac{r-\sqrt{c^2+d^2/8}}{r_0}}.
\end{equation}
In the expression above, $t_0$ is the largest interlayer hopping amplitude between $a$ and $p$ orbitals and $c$ is the interlayer distance. We choose~\cite{BianLusPRB1996} $c=3.35$\,\AA  (note that the unit cell of the bulk BSCCO contains eight CuO$_2$ layers). The parameter $r_0$ defines how fast the function $V_{\sigma}(r)$ decays with the distance between orbitals (we choose $r_0/d=0.19$). Following Ref.~\onlinecite{LDAenergyBands1995}, we take $t_0=0.27$\,eV. For the hopping amplitude  between $a$ orbitals in different layers, we have
\begin{eqnarray}
t^{aa}_{\bot\mathbf{nm}}=V_{0}(\mathbf{r}^{d}_{2\mathbf{n}}
-\mathbf{r}^{d}_{1\mathbf{m}}),\;\;V_{0}(\mathbf{r})=
t^{aa}_0e^{-\frac{r-c}{r^{aa}_{0}}}\,,
\end{eqnarray}
where we choose $t^{aa}_0=0.75$\,eV and $r^{aa}_0/d=0.3$.

Following the Slater--Koster formalism~\cite{SlaterKoster}, we tried also a bit complicated parametrization of interlayer hopping amplitudes,
which corresponds to $a$ orbital rather of $3z^2-r^2$ symmetry. Again, the functions similar to $V_{\sigma}(r)$ and $V_{0}(r)$, contain
factors depending on directional cosines and the factors describing exponential decay with the distance between ions, but now the hopping  amplitudes contain larger number of such functions. Details of this parametrization can be found elsewhere. Here, we present the results  corresponding to the first parametrization, but let us notice again that qualitatively the results are independent of the type of the hopping amplitude parametrization.

\section{Results} \label{results}

\begin{figure}[t]
\centering
\includegraphics[width=0.99\columnwidth]{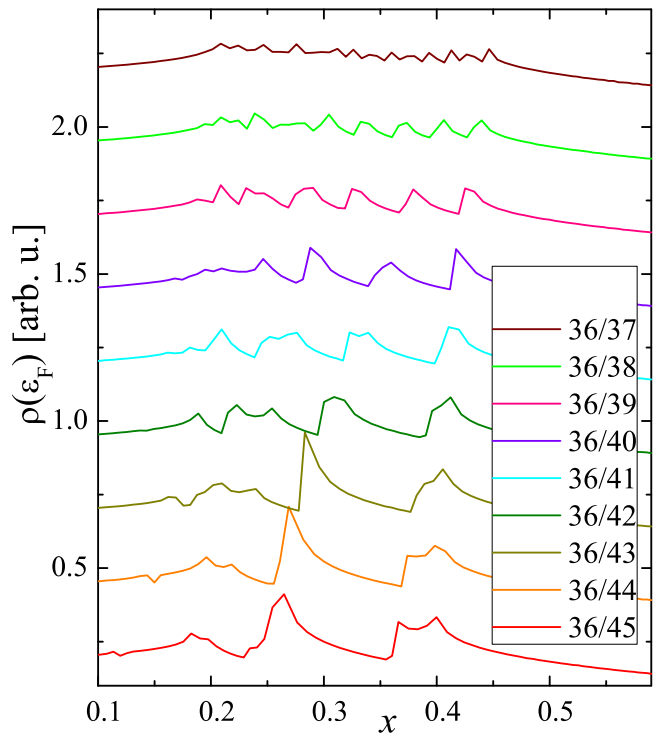}
\caption{Densities of states at Fermi level as functions of doping $x$ calculated for nine superstructures with $N_1=36$ and $N_2$ varying
form $37$ to $45$. Model parameters correspond to Fig.~\ref{FigDOStsp} with $t_{2p} = 1.5$ eV. In order to avoid logarithmic singularities,
the densities of states are averaged over the energy range $\Delta E=7$\,meV.}\label{FigDOS36_Xdoping}
\end{figure}

\begin{figure}[t]
\centering
\includegraphics[width=0.99\columnwidth]{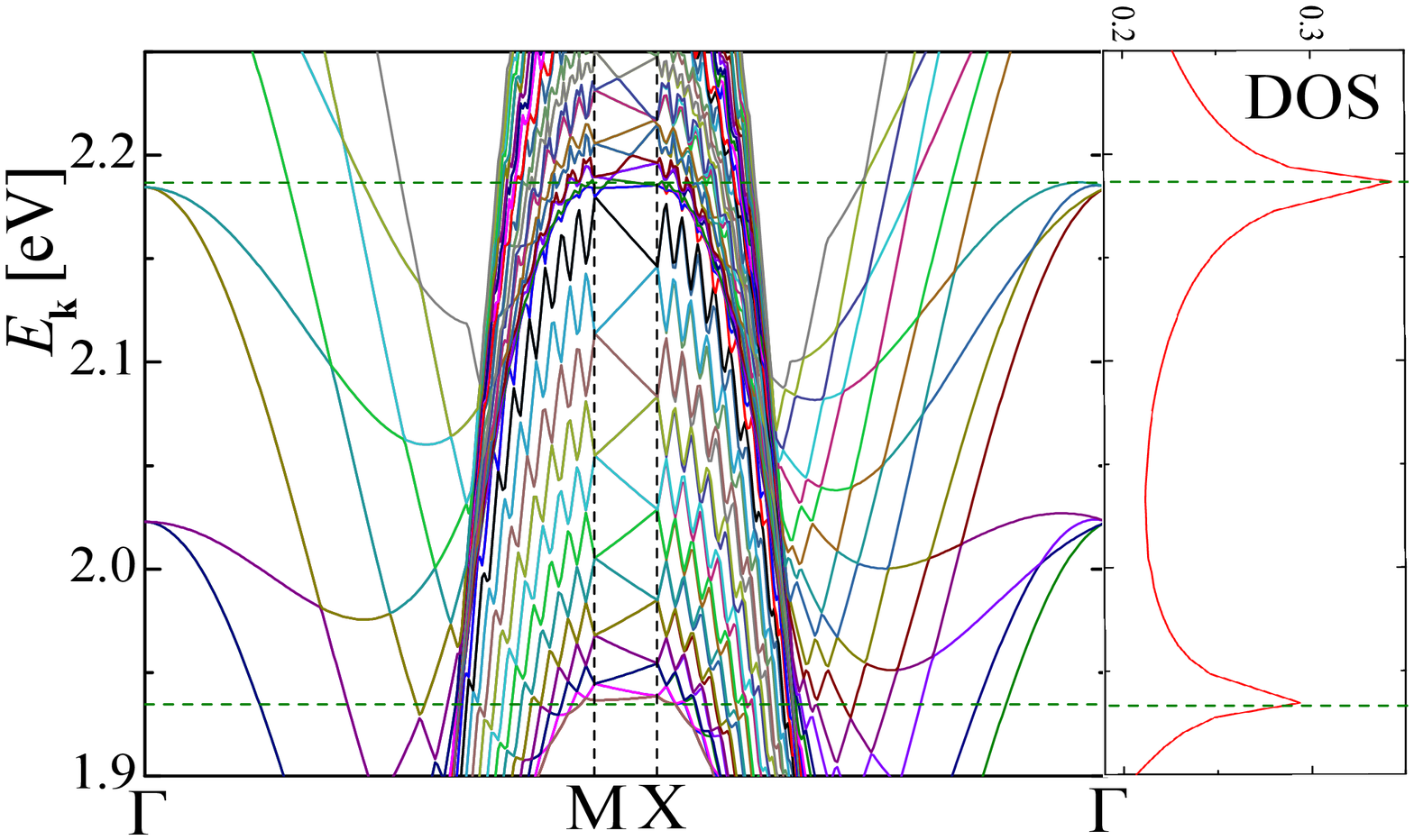}\\
\includegraphics[width=0.99\columnwidth]{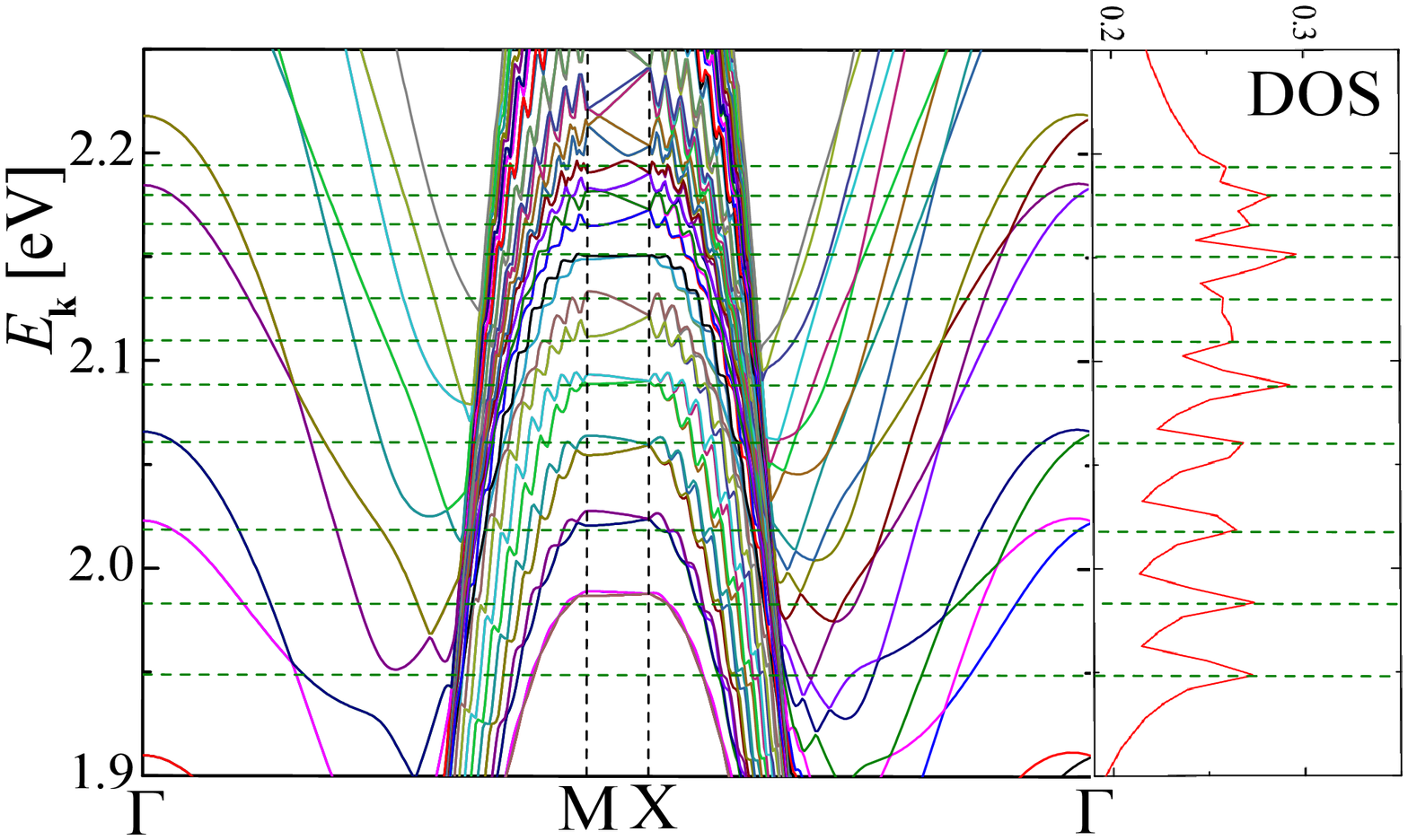}\\
\includegraphics[width=0.99\columnwidth]{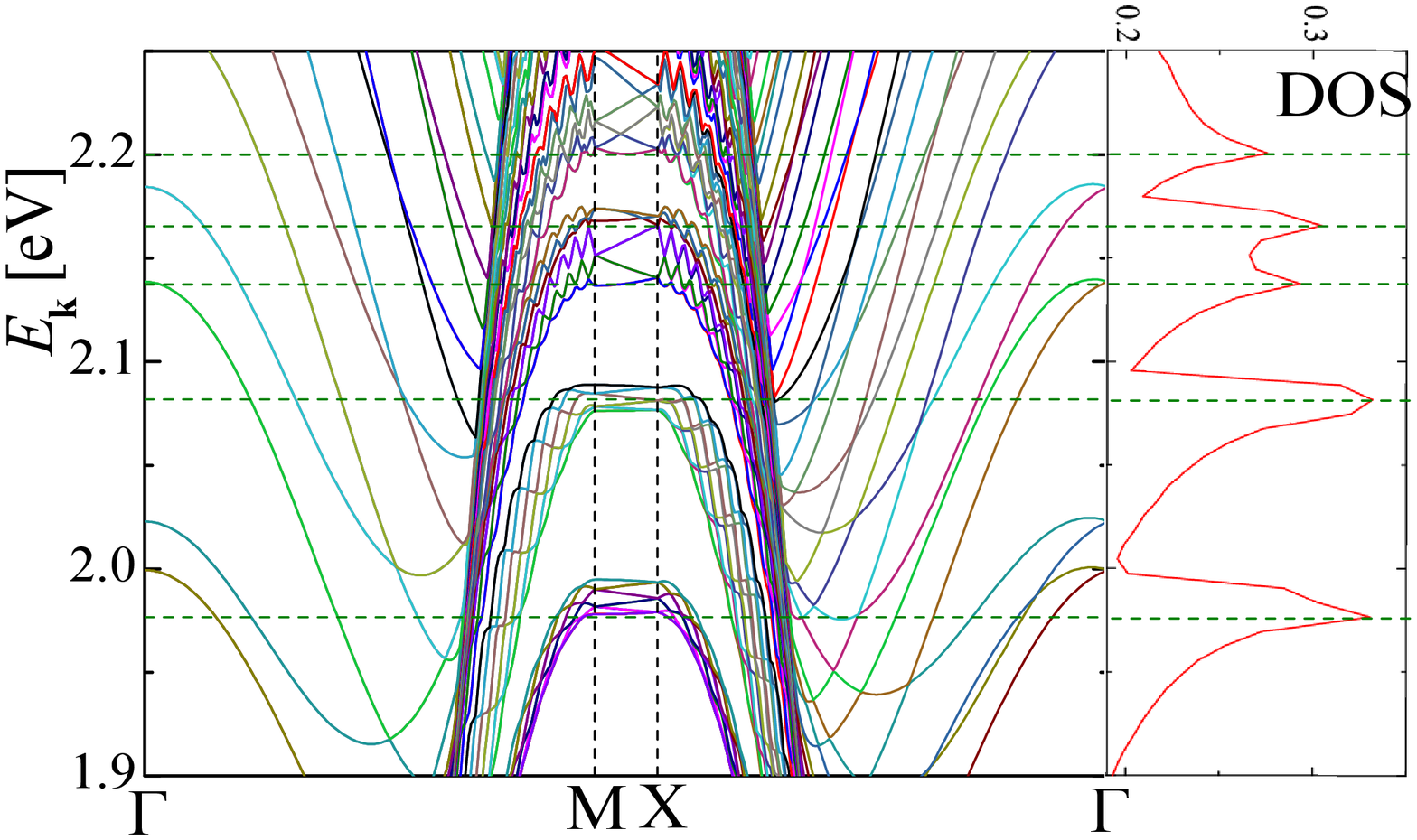}\\
\caption{Energy spectra calculated for superstructures with $N_1=36$ and $N_2=36$ (top panel, no mismatch), $N_2=38$ (middle panel), and $N_2=42$ (bottom panel). The spectra are calculated along the contour shown in Fig.~\ref{FigLayers}. The distance between $M$ and $X$ points is enhanced by factor $5$ for better view. Model parameters correspond to Fig.~\ref{FigDOStsp} with $t_{2p} = 1.5$\,eV.} \label{FigSpecDOS36_X}
\end{figure}

Here, we demonstrate the evolution of the density of states and energy spectra for bilayers at different values of the lattice mismatch.
We consider the superstructures with fixed $N_1=36$ and different $N_2$ ranging from $N_2=36$ (no mismatch) to $N_2=45$ (compressive strain).
We suggest that apical orbital $a$ is rather of $s$ orbital of copper, so its energy level lies above $x^2-y^2$ orbital, and $\varepsilon_a=6.5$\,eV.
In this case, the undoped bilayer has $5$ electrons inside elementary unit cell of each layer (four electrons on $p_{x,y}$ orbitals plus one electron
 on $d_{x^2-y^2}$ orbital). We are interested in the hole doped bilayers with doping $0<x=5-n<1$, where $n$ is the number of electrons per CuO$_2$ unit cell. Let us note here that $x$ differs from the oxygen content $y$ in the chemical formula of Bi$_2$Sr$_2$CaCu$_2$O$_{8+y}$ compound. For example, according to  Ref.~\onlinecite{PocciaPRB2011}, we have $x = y - 0.05$ for some $y$.  In the interesting energy range, the density of states of the individual layer has one van Hove singularity. For unmismatched layers, the interlayer hopping terms split this van Hove singularity into two. Their energy positions depend on the model parameters. Thus, increasing the hopping amplitude between $a$ and $p_{x,y}$ orbitals, $t_{2p}$, shifts these peaks to larger energies (smaller doping) in agreement with Ref.~\onlinecite{PavariniPRL2001}. For example, in Fig.~\ref{FigDOStsp}, we show the densities of states at Fermi level as functions of doping $x$ calculated for three different values of $t_{2p}$. Below, we consider the case of $t_{2p}=1.5$\,eV as more realistic from the viewpoint of the positions of van Hove singularities in real compounds.

\begin{figure}[t]
\centering
\vspace{2mm}
\includegraphics[width=0.99\columnwidth]{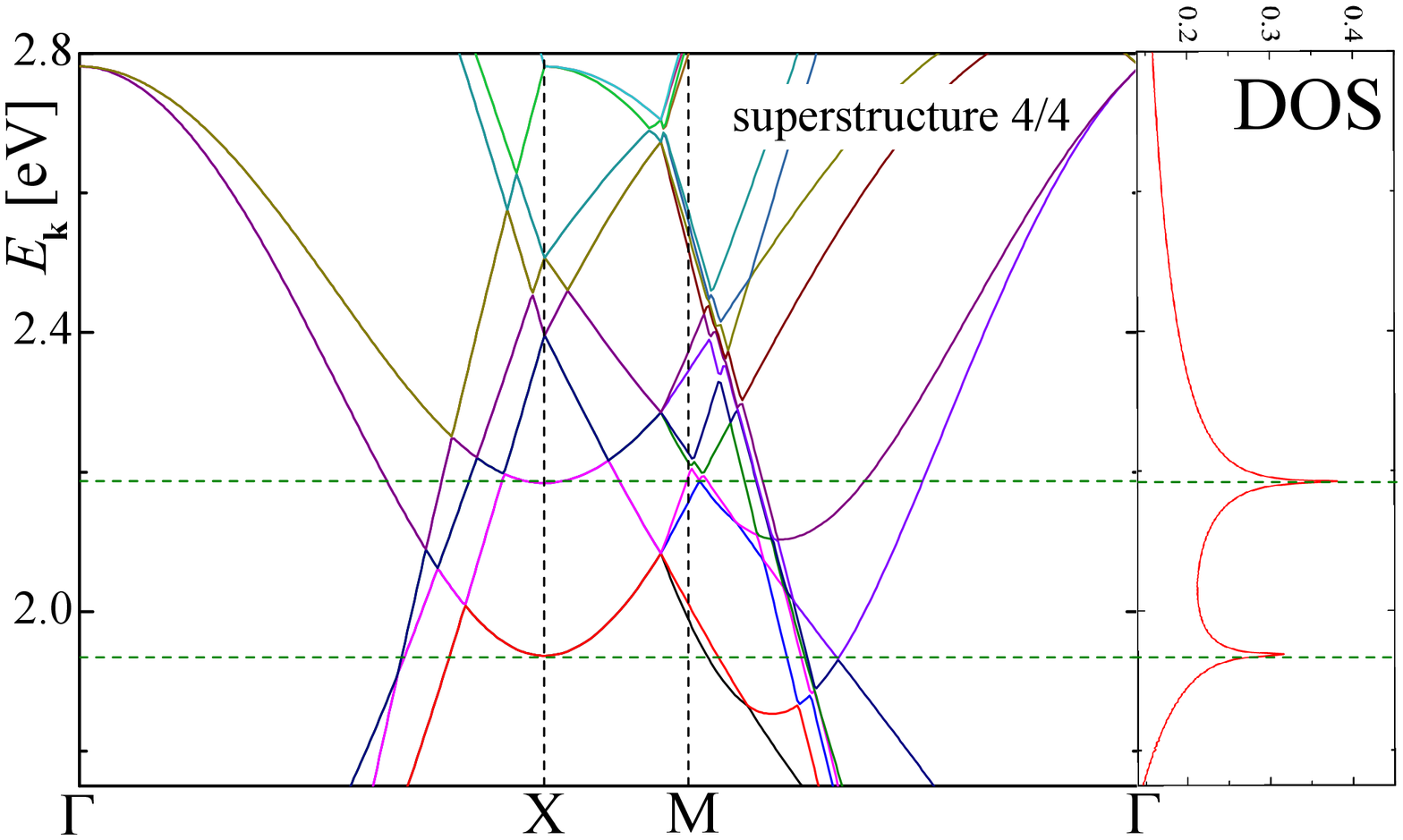}\\
\includegraphics[width=0.99\columnwidth]{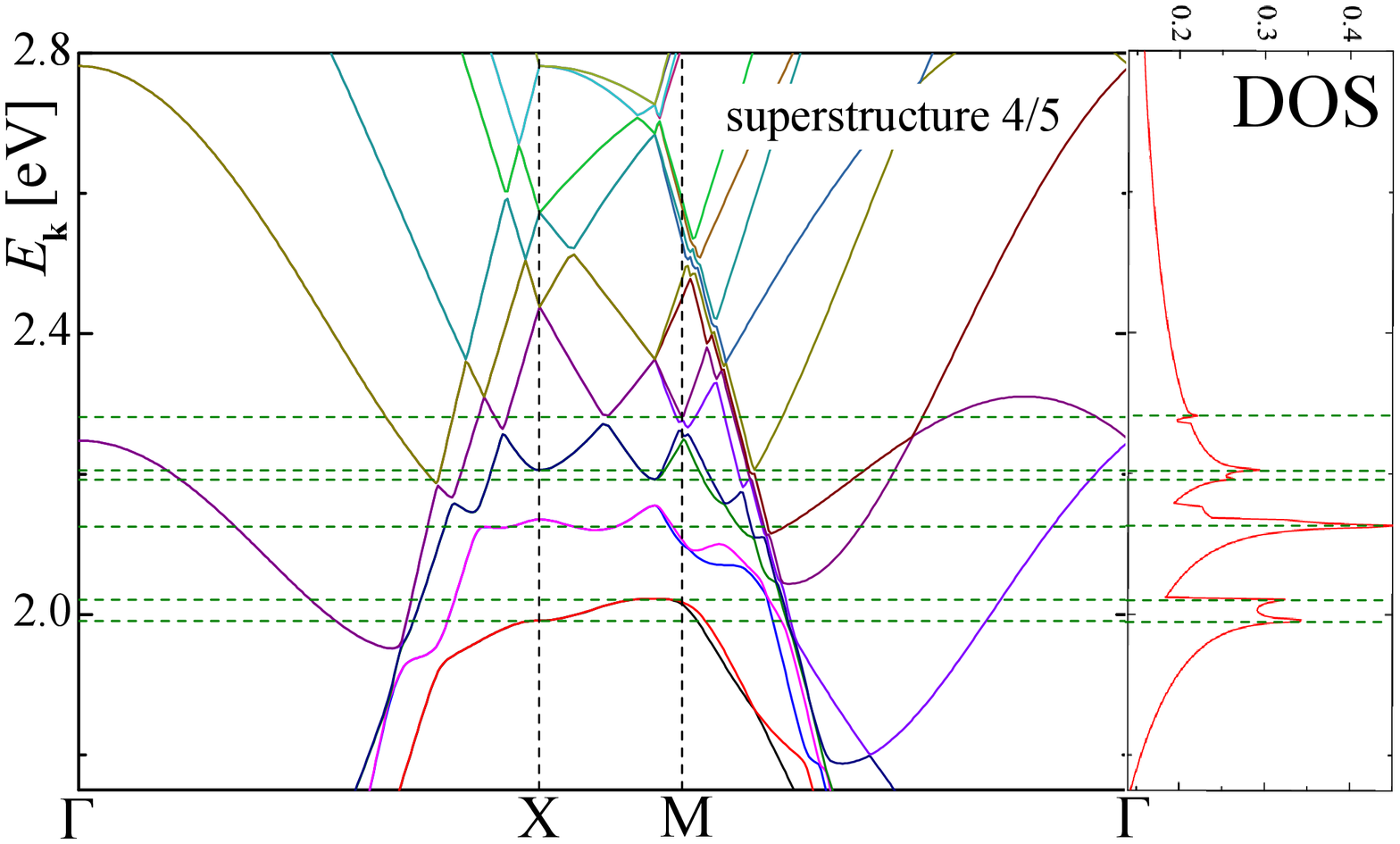}\\
\includegraphics[width=0.99\columnwidth]{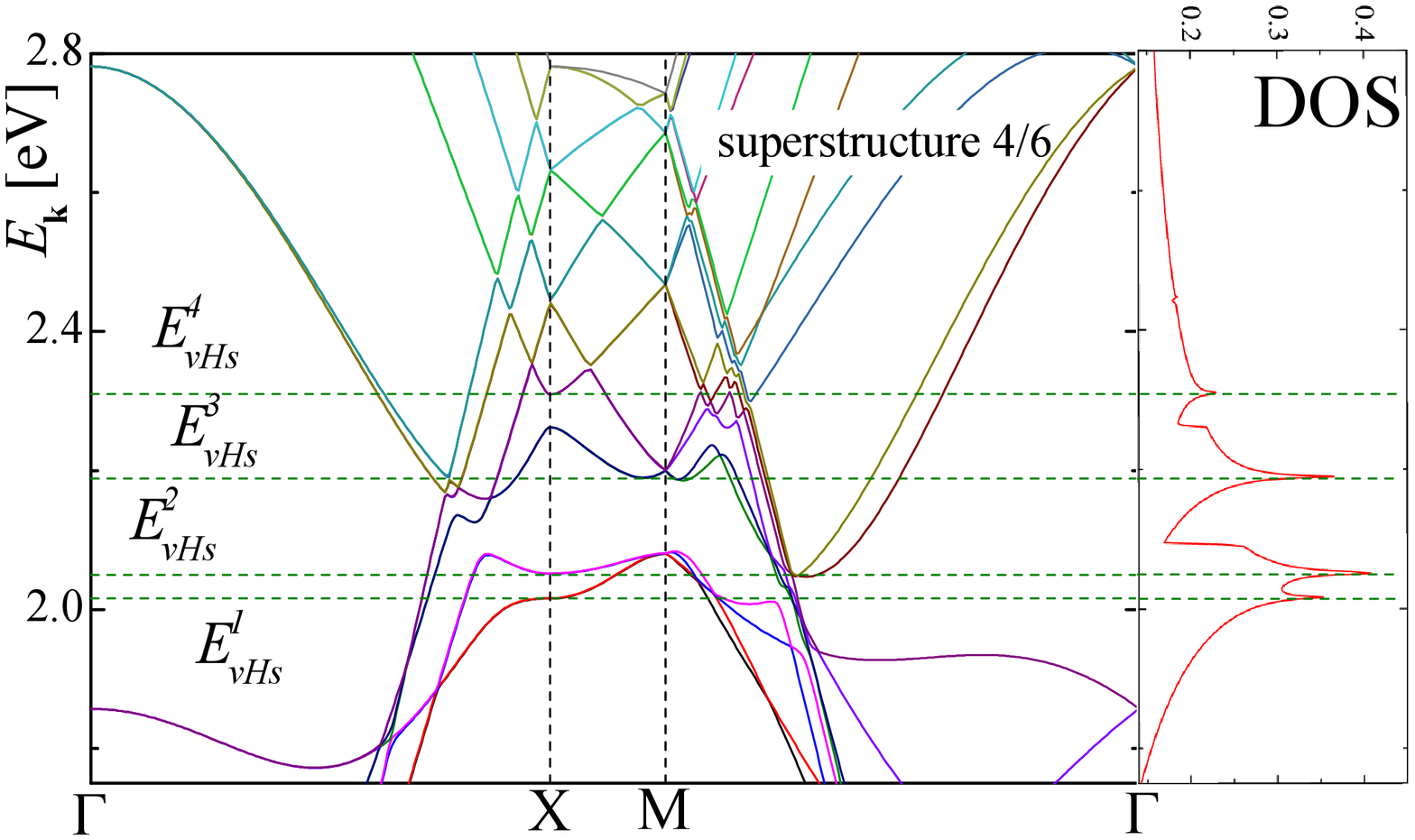}\\
\caption{Energy spectra calculated for superstructures with $N_1=4$ and $N_2=4$ (top panel, no mismatch), $N_2=5$ (middle panel), and $N_2=6$ (bottom panel). The spectra are calculated along the contour shown in Fig.~\ref{FigLayers}. Model parameters correspond to Fig.~\ref{FigDOStsp} with $t_{2p} = 1.5$ eV.} \label{FigSpecDOS4_X}
\end{figure}

The main message of our work is that the mismatch between two CuO$_2$ layers generates extra van Hove singularities. For example, in Fig.~\ref{FigDOS36_X}, we show the densities of states as function of energy calculated for nine superstructures with $N_1=36$ and $N_2$ varying form $37$ to $45$. We clearly see extra van Hove singularities approximately located between to peaks corresponding to unmismatched layers. It is interesting, that the number of these extra van Hove singularities exhibits a
counterintuitive behavior with the change of $N_2$: the number of peaks increases when $N_2$  decreases from $N_2=45$ to $N_2=37$. For better comparison, we show in a single plot, Fig.~\ref{FigDOS36_Xdoping}, the densities of states at Fermi level as functions of doping calculated for all nine superstructures with different $N_2$ and fixed $N_1=36$.

For better understanding of the origin of the extra van Hove singularities, we calculated the energy spectra for several superstructures (see Figs.~\ref{FigSpecDOS36_X} and \ref{FigSpecDOS4_X}). The spectra are calculated along the triangular contour connecting the most symmetrical points $\Gamma$, $M$, and $X$ of the reduced Brillouin zone, see the right panel of Fig.~\ref{FigLayers}. For Fig.~\ref{FigSpecDOS36_X}, there are a lot of bands (about $100$) inside the energy range shown, $1.9$\,eV$<E<2.25$\,eV. This energy range includes two van Hove singularities of unmismatched layers (upper panel of Fig.~\ref{FigSpecDOS36_X}), located at $E^1_{vHs}\cong1.93$\,eV and $E^2_{vHs}\cong2.19$\,eV. The analysis shows that the origin of these two van Hove singularities are two saddle points of certain bands located at $M$ point of the reduced Brillouin zone. In the energy range approximately between $E^1_{vHs}$ and $E^2_{vHs}$, the band crossing occurs in the region near the line connecting $M$ and $X$ points of the RBZ, as it can be seen in top panel of Fig.~\ref{FigSpecDOS36_X}. Such a picture is observed for unmismatched layers. When the layer 2 becomes stretched or compressed in comparison to layer 1, we observe a band flattening and band splitting in the region near the line connecting $M$ and $X$ points as it is seen in middle and bottom panels in Fig.~\ref{FigSpecDOS36_X}. This band flattening gives rise to extra van Hove singularities with energies lying approximately between $E^1_{vHs}$ and $E^2_{vHs}$. Situation here is very similar to that observed in twisted bilayer graphene~\cite{RozhkovPhRep2016}, where the band flattening is observed for large superstructures, when the twist angle $\theta$ decreases down to critical value $\theta_c\sim1^{\circ}$, while the moir{\'{e}} period increases as $L\propto1/\theta$. In our case, however, the band flattening occurs only in the $k_x$-direction, since the band folding is performed only in this direction in the momentum space. As we can see from Fig.~\ref{FigSpecDOS36_X}, the bands remain dispersive in the $k_y$-direction.

\begin{figure}[t]
\centering
\includegraphics[width=0.99\columnwidth]{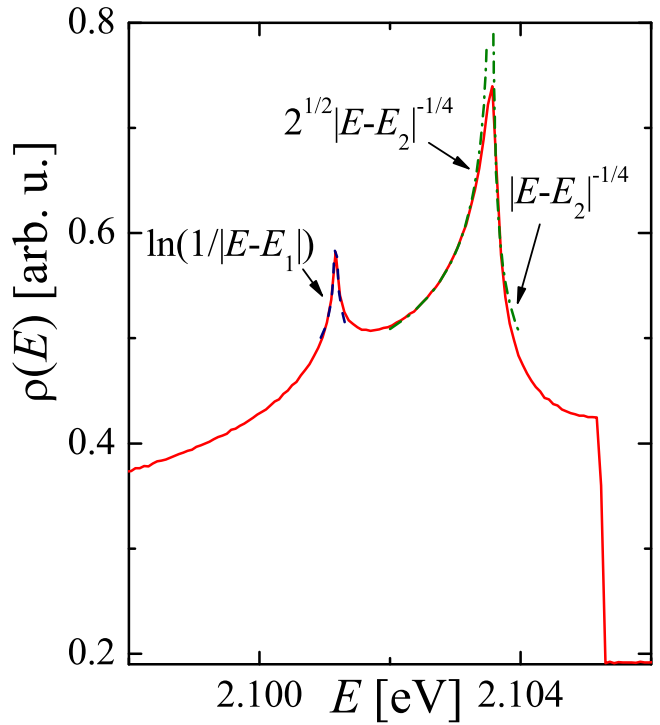}
\caption{ The largest DOS peak for the 36/43 structure. At a higher resolution, we can see that this peak splits into two ones. One of them (at $E_1$) has the logarithmic divergence, whereas another one (at $E_2$) exhibits the power-law divergence ($E^{-1/4}$) implying the existence of a higher-order VHS. Different prefactors of the power-law divergence and the exponent of the power law agree with the predictions of Refs.~\onlinecite{YuanNatCom2019} and \onlinecite{IsobePRR2019}. The stepwise change in the DOS corresponds to the band edge.}\label{asymptotics}
\end{figure}

Note that the recent studies of 2D systems and moir\'e superlattices (e.g., in twisted bilayer graphene) demonstrate that such systems often exhibit higher-order VHS characterized by the power-law divergence in the density of states in contrast to the logarithmic divergence for conventional VHS \cite{YuanNatCom2019,IsobePRR2019,ClassenPRB2019,GuerciPRR2022}. Near such VHS the dispersion is flatter than near conventional ones. To illustrate such situation in our case, we show in Fig.~\ref{asymptotics} a zoomed version of the highest DOS peak for the 36/43 structure. At a higher resolution, we can see that this peak splits into two ones. One of them has the logarithmic divergence, whereas another one exhibits the power-law divergence ($E^{-1/4}$) implying the existence of a higher-order VHS.

The spectra shown in Fig.~\ref{FigSpecDOS36_X} contain a lot of bands. For better illustration, we calculated also spectra for much smaller superlattices. Namely, we calculated spectra for superstructures with $N_1=4$ and $N_2=4,\,5$, and $6$, see Fig.~\ref{FigSpecDOS4_X}. Since, in these cases, the superlattice cell contains much smaller number of atoms, the energy range of interest (approximately between $E^1_{vHs}$ and $E^2_{vHs}$) contains much smaller number of bands. For superstructure with $N_1=4$ and $N_2=4$ (no mismatch, top panel of Fig.~\ref{FigSpecDOS4_X}), we see two parabolic bands with minima at $M$ points producing two van Hove singularities at energies $E^1_{vHs}$ and $E^2_{vHs}$. By calculating the spectrum around small circle centered at $M$ point, we have checked that $M$ point is indeed a saddle point for corresponding bands. In the momentum range near the line connecting $M$ and $X$ points, we see several band crossings. When we compress layer $2$ (superstructures with $N_2=5$ and $6$, middle and bottom panels in Fig.~\ref{FigSpecDOS4_X}), this produces the band splittings, which give rise to extra van Hove singularities.

\section{Conclusions}

In conclusion, we have studied the model of double-layer BSCCO cuprate superconductor with mismatched CuO$_2$ bilayers in diagonal direction.
The mismatch between two layers produces the superstructure in the system. This gives rise to several extra van Hove singularities
located approximately between energies $E^1_{vHs}$ and $E^2_{vHs}$ corresponding to two van Hove singularities of unmismatched layers. The number of these peaks and their energy positions depends on the quasi-commensurate superstructures of the {\it devil's staircase} in strained bilayers.

We have found that these extra van Hove singularities are closely related to the flattening and splitting of the bands inside the reduced Brillouin zone of mismatched layers. The effect of the band flattening is similar to that observed in twisted bilayer graphene~\cite{RozhkovPhRep2016}, where the large Fermi velocity reduction occurs for superstructures with small twist angles~\cite{dSPRB,ourTBLGPRB2015}. In our case, however, the band flattening appears not in the whole Brillouin zone, but only in the direction of the band folding ($k_x$-direction). In the perpendicular direction, the bands remain dispersive.

Note here that our approach relies on several approximations. For example, (i) only nearest neighbor hoppings are considered; (ii) the rotation of the orbitals in the stretched layer is not considered, and (iii) the hopping parameters in the stretched layer are assumed to remain the same as those of the non-stretched layer. Of course, these approximations make the model rather oversimplified and the extension of this model should indeed make it more realistic. However, just the structure of our calculations implies that the extra van Hove singularities will not disappear and a certain partial band flattening should remain, but maybe in some modified manner. Indeed, our additional calculations performed using different values of the hopping  integrals in each layer lead to the qualitatively similar results.

Let us also mention that DOS at the Fermi level governs to a large extent various thermodynamic properties such as charge compressibility, spin susceptibility, and specific heat. The well-known effect of the large DOS at the Fermi level on superconductivity can be here additionally enhanced since the large DOS may result in a strong screening of the repulsive interaction. This effect is the most clearly pronounced at the van Hove singularities, especially at strongly divergent high-order ones.

The present results show that BSCCO is an intrinsic multi-band system, where at high doping, the chemical potential is close to a Lifshitz transition appearing at VHS at the metal to superconductor transition (see also Ref.~\onlinecite{MazziottiPRB2021}). Therefore, superconductivity in BSCCO is not a single-band unconventional superconducting phase but a multi-gap superconductor at a Fano--Feshbach resonance between gaps in the BCS regime and in the BCS--BEC crossover as described by Perali {\it et al.} in cuprates~\cite{perali1996gap,PeraliSuST2012,chen2012superconducting,
salasnich2019screening} and appearing in diborides~\cite{agrestini2004substitution}, oxide interfaces ~\cite{bianconi2014shape} and in room temperature hydride superconductors ~\cite{mazziotti2021resonant,MazziottiEPL2021}.

Our results are supported by the recent evidence by a high resolution STM experiment of nanoscale phase separation in a  BSCCO sample doped up to $0.27$ holes/Cu site ~\cite{tromp2022puddle}, reaching the high doping range of the high order VHS singularities, which was predicted by theory to appear where the chemical potential approaches a high order VHS singularity near a Lifshitz transition in strongly correlated multiband systems ~\cite{KaganPhRep2021,RakhmanovJETPL2017,KugelPRB2008,BianconiSST2015,KugelSST2008}.
The position of the Fermi level with respect to the high order VHS peaks can be controlled by the gate voltage using modern technologies. Therefore, this makes mismatched bilayer cuprates to be an efficient system for fine tuning of the peaks of the electron density of states with respect to the Fermi level. Note that this effect is the most clearly pronounced at rather small lattice mismatch, which produces numerous additional peaks in the density of states.

\section*{Acknowledgments}
A.O.S. acknowledge the support of the Russian Science Foundation (project No. 22-22-00464) in the part concerning numerical calculations. K.I.K. acknowledge the support of the Russian Science Foundation (project No. 20-62-46047) in the part concerning the data analysis.
We are grateful to the Joint Supercomputer Center of the Russian Academy of Sciences (JSCC RAS) for the provided computational resources. A.B. acknowledges the support of the Ministry of Science and Higher Education of the Russian Federation (Agreement No. 075-15-2021-1352).

%\bibliographystyle{apsrevlong_no_issn_url}
%\bibliography{cupratesCM}

\begin{thebibliography}{74}
\expandafter\ifx\csname natexlab\endcsname\relax\def\natexlab#1{#1}\fi
\expandafter\ifx\csname bibnamefont\endcsname\relax
  \def\bibnamefont#1{#1}\fi
\expandafter\ifx\csname bibfnamefont\endcsname\relax
  \def\bibfnamefont#1{#1}\fi
\expandafter\ifx\csname citenamefont\endcsname\relax
  \def\citenamefont#1{#1}\fi

\bibitem[{\citenamefont{Liu et~al.}(2016)\citenamefont{Liu, Weiss, Duan, Cheng,
  Huang, and Duan}}]{LiuNatRevMat2016}
\bibinfo{author}{\bibfnamefont{Y.}~\bibnamefont{Liu}},
  \bibinfo{author}{\bibfnamefont{N.~O.} \bibnamefont{Weiss}},
  \bibinfo{author}{\bibfnamefont{X.}~\bibnamefont{Duan}},
  \bibinfo{author}{\bibfnamefont{H.-C.} \bibnamefont{Cheng}},
  \bibinfo{author}{\bibfnamefont{Y.}~\bibnamefont{Huang}}, \bibnamefont{and}
  \bibinfo{author}{\bibfnamefont{X.}~\bibnamefont{Duan}},
  {``}\bibinfo{title}{Van der {Waals} heterostructures and devices},{''}
  \bibinfo{journal}{Nat. Rev. Mater.} \textbf{\bibinfo{volume}{1}},
  \bibinfo{pages}{16042} (\bibinfo{year}{2016}).

\bibitem[{\citenamefont{Rozhkov et~al.}(2016)\citenamefont{Rozhkov, Sboychakov,
  Rakhmanov, and Nori}}]{RozhkovPhRep2016}
\bibinfo{author}{\bibfnamefont{A.}~\bibnamefont{Rozhkov}},
  \bibinfo{author}{\bibfnamefont{A.}~\bibnamefont{Sboychakov}},
  \bibinfo{author}{\bibfnamefont{A.}~\bibnamefont{Rakhmanov}},
  \bibnamefont{and} \bibinfo{author}{\bibfnamefont{F.}~\bibnamefont{Nori}},
  {``}\bibinfo{title}{Electronic properties of graphene-based bilayer
  systems},{''} \bibinfo{journal}{Phys. Rep.} \textbf{\bibinfo{volume}{648}},
  \bibinfo{pages}{1} (\bibinfo{year}{2016}).

\bibitem[{\citenamefont{Weston et~al.}(2020)\citenamefont{Weston, Zou,
  Enaldiev, Summerfield, Clark, Z\'olyomi, Graham, Yelgel, Magorrian, Zhou
  et~al.}}]{WestonNatNTech2020}
\bibinfo{author}{\bibfnamefont{A.}~\bibnamefont{Weston}},
  \bibinfo{author}{\bibfnamefont{Y.}~\bibnamefont{Zou}},
  \bibinfo{author}{\bibfnamefont{V.}~\bibnamefont{Enaldiev}},
  \bibinfo{author}{\bibfnamefont{A.}~\bibnamefont{Summerfield}},
  \bibinfo{author}{\bibfnamefont{N.}~\bibnamefont{Clark}},
  \bibinfo{author}{\bibfnamefont{V.}~\bibnamefont{Z\'olyomi}},
  \bibinfo{author}{\bibfnamefont{A.}~\bibnamefont{Graham}},
  \bibinfo{author}{\bibfnamefont{C.}~\bibnamefont{Yelgel}},
  \bibinfo{author}{\bibfnamefont{S.}~\bibnamefont{Magorrian}},
  \bibinfo{author}{\bibfnamefont{M.}~\bibnamefont{Zhou}}, \bibnamefont{et~al.},
  {``}\bibinfo{title}{Atomic reconstruction in twisted bilayers of transition
  metal dichalcogenides},{''} \bibinfo{journal}{Nat. Nanotechnol.}
  \textbf{\bibinfo{volume}{15}}, \bibinfo{pages}{592} (\bibinfo{year}{2020}).

\bibitem[{\citenamefont{Zhao et~al.}(2021)\citenamefont{Zhao, Zhu, Nie, Li,
  Wang, Dou, Hu, Xian, Meng, and Li}}]{ZhaoNatMat2021}
\bibinfo{author}{\bibfnamefont{W.-M.} \bibnamefont{Zhao}},
  \bibinfo{author}{\bibfnamefont{L.}~\bibnamefont{Zhu}},
  \bibinfo{author}{\bibfnamefont{Z.}~\bibnamefont{Nie}},
  \bibinfo{author}{\bibfnamefont{Q.-Y.} \bibnamefont{Li}},
  \bibinfo{author}{\bibfnamefont{Q.-W.} \bibnamefont{Wang}},
  \bibinfo{author}{\bibfnamefont{L.-G.} \bibnamefont{Dou}},
  \bibinfo{author}{\bibfnamefont{J.-G.} \bibnamefont{Hu}},
  \bibinfo{author}{\bibfnamefont{L.}~\bibnamefont{Xian}},
  \bibinfo{author}{\bibfnamefont{S.}~\bibnamefont{Meng}}, \bibnamefont{and}
  \bibinfo{author}{\bibfnamefont{S.-C.} \bibnamefont{Li}},
  {``}\bibinfo{title}{Moir\'{e} enhanced charge density wave state in twisted
  {1T-TiTe$_2$/1T-TiSe$_2$} heterostructures},{''} \bibinfo{journal}{Nat.
  Mater.} \textbf{\bibinfo{volume}{21}}, \bibinfo{pages}{284}
  (\bibinfo{year}{2021}).

\bibitem[{\citenamefont{Tong et~al.}(2017)\citenamefont{Tong, Yu, Zhu, Wang,
  Xu, and Yao}}]{TongNatPh2017}
\bibinfo{author}{\bibfnamefont{Q.}~\bibnamefont{Tong}},
  \bibinfo{author}{\bibfnamefont{H.}~\bibnamefont{Yu}},
  \bibinfo{author}{\bibfnamefont{Q.}~\bibnamefont{Zhu}},
  \bibinfo{author}{\bibfnamefont{Y.}~\bibnamefont{Wang}},
  \bibinfo{author}{\bibfnamefont{X.}~\bibnamefont{Xu}}, \bibnamefont{and}
  \bibinfo{author}{\bibfnamefont{W.}~\bibnamefont{Yao}},
  {``}\bibinfo{title}{Topological mosaics in moir\'{e} superlattices of van der
  {Waals} heterobilayers},{''} \bibinfo{journal}{Nat. Phys.}
  \textbf{\bibinfo{volume}{13}}, \bibinfo{pages}{356} (\bibinfo{year}{2017}).

\bibitem[{\citenamefont{Ren et~al.}(2020)\citenamefont{Ren, Zhang, Liu, and
  He}}]{RenChinPh2020}
\bibinfo{author}{\bibfnamefont{Y.-N.} \bibnamefont{Ren}},
  \bibinfo{author}{\bibfnamefont{Y.}~\bibnamefont{Zhang}},
  \bibinfo{author}{\bibfnamefont{Y.-W.} \bibnamefont{Liu}}, \bibnamefont{and}
  \bibinfo{author}{\bibfnamefont{L.}~\bibnamefont{He}},
  {``}\bibinfo{title}{Twistronics in graphene-based van der {Waals}
  structures},{''} \bibinfo{journal}{Chin. Phys. B}
  \textbf{\bibinfo{volume}{29}}, \bibinfo{pages}{117303}
  (\bibinfo{year}{2020}).

\bibitem[{\citenamefont{Krockenberger et~al.}(2021)\citenamefont{Krockenberger,
  Ikeda, and Yamamoto}}]{KrockenbergerACSOmega2021}
\bibinfo{author}{\bibfnamefont{Y.}~\bibnamefont{Krockenberger}},
  \bibinfo{author}{\bibfnamefont{A.}~\bibnamefont{Ikeda}}, \bibnamefont{and}
  \bibinfo{author}{\bibfnamefont{H.}~\bibnamefont{Yamamoto}},
  {``}\bibinfo{title}{Atomic stripe formation in infinite-layer cuprates},{''}
  \bibinfo{journal}{ACS Omega} \textbf{\bibinfo{volume}{6}},
  \bibinfo{pages}{21884} (\bibinfo{year}{2021}).

\bibitem[{\citenamefont{Park et~al.}(2021)\citenamefont{Park, Oh, Song, and
  Kang}}]{ParkProgSCCr2021}
\bibinfo{author}{\bibfnamefont{H.}~\bibnamefont{Park}},
  \bibinfo{author}{\bibfnamefont{J.}~\bibnamefont{Oh}},
  \bibinfo{author}{\bibfnamefont{B.}~\bibnamefont{Song}}, \bibnamefont{and}
  \bibinfo{author}{\bibfnamefont{B.}~\bibnamefont{Kang}},
  {``}\bibinfo{title}{Structural suitability of {GdFeO$_3$} as a magnetic
  buffer layer for {GdBa$_2$Cu$_3$O$_{7-x}$} superconducting thin films},{''}
  \bibinfo{journal}{Progr. Supercond. Cryog.} \textbf{\bibinfo{volume}{23}},
  \bibinfo{pages}{14} (\bibinfo{year}{2021}).

\bibitem[{\citenamefont{Conradson et~al.}(2020)\citenamefont{Conradson,
  Geballe, Jin, Cao, Gauzzi, Karppinen, Baldinozzi, Li, Gilioli, Jiang
  et~al.}}]{ConradsonPNAS2020}
\bibinfo{author}{\bibfnamefont{S.~D.} \bibnamefont{Conradson}},
  \bibinfo{author}{\bibfnamefont{T.~H.} \bibnamefont{Geballe}},
  \bibinfo{author}{\bibfnamefont{C.-Q.} \bibnamefont{Jin}},
  \bibinfo{author}{\bibfnamefont{L.-P.} \bibnamefont{Cao}},
  \bibinfo{author}{\bibfnamefont{A.}~\bibnamefont{Gauzzi}},
  \bibinfo{author}{\bibfnamefont{M.}~\bibnamefont{Karppinen}},
  \bibinfo{author}{\bibfnamefont{G.}~\bibnamefont{Baldinozzi}},
  \bibinfo{author}{\bibfnamefont{W.-M.} \bibnamefont{Li}},
  \bibinfo{author}{\bibfnamefont{E.}~\bibnamefont{Gilioli}},
  \bibinfo{author}{\bibfnamefont{J.~M.} \bibnamefont{Jiang}},
  \bibnamefont{et~al.}, {``}\bibinfo{title}{Nonadiabatic coupling of the
  dynamical structure to the superconductivity in
  {YSr$_2$Cu$_{2.75}$Mo$_{0.25}$O$_{7.54}$} and {Sr$_2$CuO$_{3.3}$}},{''}
  \bibinfo{journal}{PNAS} \textbf{\bibinfo{volume}{117}},
  \bibinfo{pages}{33099} (\bibinfo{year}{2020}).

\bibitem[{\citenamefont{Sederholm et~al.}(2021)\citenamefont{Sederholm,
  Conradson, Geballe, Jin, Gauzzi, Gilioli, Karppinen, and
  Baldinozzi}}]{SederholmCondMat2021}
\bibinfo{author}{\bibfnamefont{L.}~\bibnamefont{Sederholm}},
  \bibinfo{author}{\bibfnamefont{S.~D.} \bibnamefont{Conradson}},
  \bibinfo{author}{\bibfnamefont{T.~H.} \bibnamefont{Geballe}},
  \bibinfo{author}{\bibfnamefont{C.-Q.} \bibnamefont{Jin}},
  \bibinfo{author}{\bibfnamefont{A.}~\bibnamefont{Gauzzi}},
  \bibinfo{author}{\bibfnamefont{E.}~\bibnamefont{Gilioli}},
  \bibinfo{author}{\bibfnamefont{M.}~\bibnamefont{Karppinen}},
  \bibnamefont{and}
  \bibinfo{author}{\bibfnamefont{G.}~\bibnamefont{Baldinozzi}},
  {``}\bibinfo{title}{Extremely overdoped superconducting cuprates via high
  pressure oxygenation methods},{''} \bibinfo{journal}{Condens. Matter}
  \textbf{\bibinfo{volume}{6}} (\bibinfo{year}{2021}).

\bibitem[{\citenamefont{Jang et~al.}(2017)\citenamefont{Jang, Asano, Fujita,
  Hashimoto, Lu, Burns, Kao, and Lee}}]{JangPRX2017}
\bibinfo{author}{\bibfnamefont{H.}~\bibnamefont{Jang}},
  \bibinfo{author}{\bibfnamefont{S.}~\bibnamefont{Asano}},
  \bibinfo{author}{\bibfnamefont{M.}~\bibnamefont{Fujita}},
  \bibinfo{author}{\bibfnamefont{M.}~\bibnamefont{Hashimoto}},
  \bibinfo{author}{\bibfnamefont{D.~H.} \bibnamefont{Lu}},
  \bibinfo{author}{\bibfnamefont{C.~A.} \bibnamefont{Burns}},
  \bibinfo{author}{\bibfnamefont{C.-C.} \bibnamefont{Kao}}, \bibnamefont{and}
  \bibinfo{author}{\bibfnamefont{J.-S.} \bibnamefont{Lee}},
  {``}\bibinfo{title}{Superconductivity-insensitive order at
  $q\ensuremath{\sim}1/4$ in electron-doped cuprates},{''}
  \bibinfo{journal}{Phys. Rev. X} \textbf{\bibinfo{volume}{7}},
  \bibinfo{pages}{041066} (\bibinfo{year}{2017}).

\bibitem[{\citenamefont{Bianconi
  et~al.}(1996{\natexlab{a}})\citenamefont{Bianconi, Saini, Lanzara, Missori,
  Rossetti, Oyanagi, Yamaguchi, Oka, and Ito}}]{BianconiPRL1996}
\bibinfo{author}{\bibfnamefont{A.}~\bibnamefont{Bianconi}},
  \bibinfo{author}{\bibfnamefont{N.~L.} \bibnamefont{Saini}},
  \bibinfo{author}{\bibfnamefont{A.}~\bibnamefont{Lanzara}},
  \bibinfo{author}{\bibfnamefont{M.}~\bibnamefont{Missori}},
  \bibinfo{author}{\bibfnamefont{T.}~\bibnamefont{Rossetti}},
  \bibinfo{author}{\bibfnamefont{H.}~\bibnamefont{Oyanagi}},
  \bibinfo{author}{\bibfnamefont{H.}~\bibnamefont{Yamaguchi}},
  \bibinfo{author}{\bibfnamefont{K.}~\bibnamefont{Oka}}, \bibnamefont{and}
  \bibinfo{author}{\bibfnamefont{T.}~\bibnamefont{Ito}},
  {``}\bibinfo{title}{Determination of the local lattice distortions in the
  {Cu${\mathrm{O}}_{2}$} plane of
  {L${\mathrm{a}}_{1.85}$S${\mathrm{r}}_{0.15}$Cu${\mathrm{O}}_{4}$}},{''}
  \bibinfo{journal}{Phys. Rev. Lett.} \textbf{\bibinfo{volume}{76}},
  \bibinfo{pages}{3412} (\bibinfo{year}{1996}{\natexlab{a}}).

\bibitem[{\citenamefont{Saini et~al.}(2003)\citenamefont{Saini, Oyanagi, Ito,
  Scagnoli, Filippi, Agrestini, Campi, Oka, and Bianconi}}]{SainiEPJB2003}
\bibinfo{author}{\bibfnamefont{N.}~\bibnamefont{Saini}},
  \bibinfo{author}{\bibfnamefont{H.}~\bibnamefont{Oyanagi}},
  \bibinfo{author}{\bibfnamefont{T.}~\bibnamefont{Ito}},
  \bibinfo{author}{\bibfnamefont{V.}~\bibnamefont{Scagnoli}},
  \bibinfo{author}{\bibfnamefont{M.}~\bibnamefont{Filippi}},
  \bibinfo{author}{\bibfnamefont{S.}~\bibnamefont{Agrestini}},
  \bibinfo{author}{\bibfnamefont{G.}~\bibnamefont{Campi}},
  \bibinfo{author}{\bibfnamefont{K.}~\bibnamefont{Oka}}, \bibnamefont{and}
  \bibinfo{author}{\bibfnamefont{A.}~\bibnamefont{Bianconi}},
  {``}\bibinfo{title}{Temperature dependent local {Cu-O} displacements from
  underdoped to overdoped {La-Sr-Cu-O} superconductor},{''}
  \bibinfo{journal}{Eur. Phys. J. B} \textbf{\bibinfo{volume}{36}},
  \bibinfo{pages}{75} (\bibinfo{year}{2003}).

\bibitem[{\citenamefont{Bianconi
  et~al.}(1996{\natexlab{b}})\citenamefont{Bianconi, Lusignoli, Saini, Bordet,
  Kvick, and Radaelli}}]{BianconiPRB1996}
\bibinfo{author}{\bibfnamefont{A.}~\bibnamefont{Bianconi}},
  \bibinfo{author}{\bibfnamefont{M.}~\bibnamefont{Lusignoli}},
  \bibinfo{author}{\bibfnamefont{N.~L.} \bibnamefont{Saini}},
  \bibinfo{author}{\bibfnamefont{P.}~\bibnamefont{Bordet}},
  \bibinfo{author}{\bibfnamefont{A.}~\bibnamefont{Kvick}}, \bibnamefont{and}
  \bibinfo{author}{\bibfnamefont{P.~G.} \bibnamefont{Radaelli}},
  {``}\bibinfo{title}{Stripe structure of the ${\mathrm{CuO}}_{2}$ plane in
  ${\mathrm{Bi}}_{2}$${\mathrm{Sr}}_{2}$${\mathrm{CaCu}}_{2}$${\mathrm{O}}_{8+\mathit{y}}$
  by anomalous x-ray diffraction},{''} \bibinfo{journal}{Phys. Rev. B}
  \textbf{\bibinfo{volume}{54}}, \bibinfo{pages}{4310}
  (\bibinfo{year}{1996}{\natexlab{b}}).

\bibitem[{\citenamefont{Bianconi
  et~al.}(1996{\natexlab{c}})\citenamefont{Bianconi, Saini, Rossetti, Lanzara,
  Perali, Missori, Oyanagi, Yamaguchi, Nishihara, and Ha}}]{BianconiPRB1996a}
\bibinfo{author}{\bibfnamefont{A.}~\bibnamefont{Bianconi}},
  \bibinfo{author}{\bibfnamefont{N.~L.} \bibnamefont{Saini}},
  \bibinfo{author}{\bibfnamefont{T.}~\bibnamefont{Rossetti}},
  \bibinfo{author}{\bibfnamefont{A.}~\bibnamefont{Lanzara}},
  \bibinfo{author}{\bibfnamefont{A.}~\bibnamefont{Perali}},
  \bibinfo{author}{\bibfnamefont{M.}~\bibnamefont{Missori}},
  \bibinfo{author}{\bibfnamefont{H.}~\bibnamefont{Oyanagi}},
  \bibinfo{author}{\bibfnamefont{H.}~\bibnamefont{Yamaguchi}},
  \bibinfo{author}{\bibfnamefont{Y.}~\bibnamefont{Nishihara}},
  \bibnamefont{and} \bibinfo{author}{\bibfnamefont{D.~H.} \bibnamefont{Ha}},
  {``}\bibinfo{title}{Stripe structure in the ${\mathrm{CuO}}_{2}$ plane of
  perovskite superconductors},{''} \bibinfo{journal}{Phys. Rev. B}
  \textbf{\bibinfo{volume}{54}}, \bibinfo{pages}{12018}
  (\bibinfo{year}{1996}{\natexlab{c}}).

\bibitem[{\citenamefont{{Di Castro} et~al.}(2000)\citenamefont{{Di Castro},
  Bianconi, Colapietro, Pifferi, Saini, Agrestini, and
  Bianconi}}]{DiCastroEPJB2000}
\bibinfo{author}{\bibfnamefont{D.}~\bibnamefont{{Di Castro}}},
  \bibinfo{author}{\bibfnamefont{G.}~\bibnamefont{Bianconi}},
  \bibinfo{author}{\bibfnamefont{M.}~\bibnamefont{Colapietro}},
  \bibinfo{author}{\bibfnamefont{A.}~\bibnamefont{Pifferi}},
  \bibinfo{author}{\bibfnamefont{N.}~\bibnamefont{Saini}},
  \bibinfo{author}{\bibfnamefont{S.}~\bibnamefont{Agrestini}},
  \bibnamefont{and} \bibinfo{author}{\bibfnamefont{A.}~\bibnamefont{Bianconi}},
  {``}\bibinfo{title}{Temperature dependent local {Cu-O} displacements from
  underdoped to overdoped {La-Sr-Cu-O} superconductor},{''}
  \bibinfo{journal}{Eur. Phys. J. B} \textbf{\bibinfo{volume}{18}},
  \bibinfo{pages}{617} (\bibinfo{year}{2000}).

\bibitem[{\citenamefont{Bianconi et~al.}(2000)\citenamefont{Bianconi, Bianconi,
  Caprara, Castro, Oyanagi, and Saini}}]{BianconiJPCM2000}
\bibinfo{author}{\bibfnamefont{A.}~\bibnamefont{Bianconi}},
  \bibinfo{author}{\bibfnamefont{G.}~\bibnamefont{Bianconi}},
  \bibinfo{author}{\bibfnamefont{S.}~\bibnamefont{Caprara}},
  \bibinfo{author}{\bibfnamefont{D.~D.} \bibnamefont{Castro}},
  \bibinfo{author}{\bibfnamefont{H.}~\bibnamefont{Oyanagi}}, \bibnamefont{and}
  \bibinfo{author}{\bibfnamefont{N.~L.} \bibnamefont{Saini}},
  {``}\bibinfo{title}{The stripe critical point for cuprates},{''}
  \bibinfo{journal}{J. Phys : Condens. Matter} \textbf{\bibinfo{volume}{12}},
  \bibinfo{pages}{10655} (\bibinfo{year}{2000}).

\bibitem[{\citenamefont{Agrestini et~al.}(2003)\citenamefont{Agrestini, Saini,
  Bianconi, and Bianconi}}]{AgrestiniJPhA2003}
\bibinfo{author}{\bibfnamefont{S.}~\bibnamefont{Agrestini}},
  \bibinfo{author}{\bibfnamefont{N.}~\bibnamefont{Saini}},
  \bibinfo{author}{\bibfnamefont{G.}~\bibnamefont{Bianconi}}, \bibnamefont{and}
  \bibinfo{author}{\bibfnamefont{A.}~\bibnamefont{Bianconi}},
  {``}\bibinfo{title}{The strain of {CuO$_2$} lattice: the second variable for
  the phase diagram of cuprate perovskites},{''} \bibinfo{journal}{J. Phys. A:
  Math. Gen.} \textbf{\bibinfo{volume}{36}}, \bibinfo{pages}{9133}
  (\bibinfo{year}{2003}).

\bibitem[{\citenamefont{Wiegers}(1996)}]{WiegersProgSSCh1996}
\bibinfo{author}{\bibfnamefont{G.}~\bibnamefont{Wiegers}},
  {``}\bibinfo{title}{Misfit layer compounds: {Structures} and physical
  properties},{''} \bibinfo{journal}{Prog. Solid State Chem.}
  \textbf{\bibinfo{volume}{24}}, \bibinfo{pages}{1} (\bibinfo{year}{1996}).

\bibitem[{\citenamefont{Janssen et~al.}(2004)\citenamefont{Janssen, Janner,
  {Looijenga-Vos}, and {de Wolff}}}]{IntTabl2004}
\bibinfo{author}{\bibfnamefont{T.}~\bibnamefont{Janssen}},
  \bibinfo{author}{\bibfnamefont{A.}~\bibnamefont{Janner}},
  \bibinfo{author}{\bibfnamefont{A.}~\bibnamefont{{Looijenga-Vos}}},
  \bibnamefont{and} \bibinfo{author}{\bibfnamefont{P.}~\bibnamefont{{de
  Wolff}}}, \emph{\bibinfo{title}{International Tables for Crystallography,
  Mathematical, Physical and Chemical Tables, E. Prince, ed.}}
  (\bibinfo{publisher}{Wiley, New York}, \bibinfo{year}{2004}),
  vol.~\bibinfo{volume}{C} of \emph{\bibinfo{series}{IUCr Series}}, chap.
  \bibinfo{chapter}{in 9.8. Incommensurate and Commensurate Modulated
  Structures}, \bibinfo{edition}{3rd} ed.

\bibitem[{\citenamefont{Etrillard et~al.}(2000)\citenamefont{Etrillard,
  Bourges, and Lin}}]{EtrillardPRB2000}
\bibinfo{author}{\bibfnamefont{J.}~\bibnamefont{Etrillard}},
  \bibinfo{author}{\bibfnamefont{P.}~\bibnamefont{Bourges}}, \bibnamefont{and}
  \bibinfo{author}{\bibfnamefont{C.~T.} \bibnamefont{Lin}},
  {``}\bibinfo{title}{Incommensurate composite structure of the superconductor
  ${\mathrm{Bi}}_{2}{\mathrm{Sr}}_{2}{\mathrm{CaCu}}_{2}{\mathrm{O}}_{8+\ensuremath{\delta}}$},{''}
  \bibinfo{journal}{Phys. Rev. B} \textbf{\bibinfo{volume}{62}},
  \bibinfo{pages}{150} (\bibinfo{year}{2000}).

\bibitem[{\citenamefont{Grebille et~al.}(2001)\citenamefont{Grebille, Leligny,
  and P\'erez}}]{GrebillePRB2001comment}
\bibinfo{author}{\bibfnamefont{D.}~\bibnamefont{Grebille}},
  \bibinfo{author}{\bibfnamefont{H.}~\bibnamefont{Leligny}}, \bibnamefont{and}
  \bibinfo{author}{\bibfnamefont{O.}~\bibnamefont{P\'erez}},
  {``}\bibinfo{title}{Comment on ``Incommensurate composite structure of the
  superconductor {Bi$_2$Sr$_2$CaCu$_2$O$_{8+\delta}$}''},{''}
  \bibinfo{journal}{Phys. Rev. B} \textbf{\bibinfo{volume}{64}},
  \bibinfo{pages}{106501} (\bibinfo{year}{2001}).

\bibitem[{\citenamefont{Bianconi}(1994)}]{BianconiSSC1994}
\bibinfo{author}{\bibfnamefont{A.}~\bibnamefont{Bianconi}},
  {``}\bibinfo{title}{On the {Fermi} liquid coupled with a generalized {Wigner}
  polaronic {CDW} giving high {T$_c$} superconductivity},{''}
  \bibinfo{journal}{Solid State Commun.} \textbf{\bibinfo{volume}{91}},
  \bibinfo{pages}{1} (\bibinfo{year}{1994}).

\bibitem[{\citenamefont{Beskrovnyy and
  Jir\'ak}(2012)}]{BeskrovnyyJPhConfSer2012}
\bibinfo{author}{\bibfnamefont{A.}~\bibnamefont{Beskrovnyy}} \bibnamefont{and}
  \bibinfo{author}{\bibfnamefont{Z.}~\bibnamefont{Jir\'ak}},
  {``}\bibinfo{title}{Structural modulation in
  {Bi$_2$Sr$_2$Ca$_{0.4}$Y$_{0.6}$Cu2O$_{8+\delta}$}},{''} \bibinfo{journal}{J.
  Phys. Conf. Ser.} \textbf{\bibinfo{volume}{340}} (\bibinfo{year}{2012}).

\bibitem[{\citenamefont{P\'erez et~al.}(1997)\citenamefont{P\'erez, Leligny,
  Grebille, Gren\^eche, Labb\'e, Groult, and Raveau}}]{PerezPRB1997}
\bibinfo{author}{\bibfnamefont{O.}~\bibnamefont{P\'erez}},
  \bibinfo{author}{\bibfnamefont{H.}~\bibnamefont{Leligny}},
  \bibinfo{author}{\bibfnamefont{D.}~\bibnamefont{Grebille}},
  \bibinfo{author}{\bibfnamefont{J.~M.} \bibnamefont{Gren\^eche}},
  \bibinfo{author}{\bibfnamefont{P.}~\bibnamefont{Labb\'e}},
  \bibinfo{author}{\bibfnamefont{D.}~\bibnamefont{Groult}}, \bibnamefont{and}
  \bibinfo{author}{\bibfnamefont{B.}~\bibnamefont{Raveau}},
  {``}\bibinfo{title}{Disorder phenomena in the incommensurate compound
  {Bi$_{2+\mathrm{x}}$Sr$_{3-\mathrm{x}}$Fe$_2$O$_{9+\delta}$}},{''}
  \bibinfo{journal}{Phys. Rev. B} \textbf{\bibinfo{volume}{55}},
  \bibinfo{pages}{1236} (\bibinfo{year}{1997}).

\bibitem[{\citenamefont{Sedykh et~al.}(2003)\citenamefont{Sedykh, Shekhtman,
  Smirnova, Bagautdinov, Suvorov, and Dubovitskii}}]{SedykhPhC2003}
\bibinfo{author}{\bibfnamefont{V.}~\bibnamefont{Sedykh}},
  \bibinfo{author}{\bibfnamefont{V.}~\bibnamefont{Shekhtman}},
  \bibinfo{author}{\bibfnamefont{I.}~\bibnamefont{Smirnova}},
  \bibinfo{author}{\bibfnamefont{B.}~\bibnamefont{Bagautdinov}},
  \bibinfo{author}{\bibfnamefont{E.}~\bibnamefont{Suvorov}}, \bibnamefont{and}
  \bibinfo{author}{\bibfnamefont{A.}~\bibnamefont{Dubovitskii}},
  {``}\bibinfo{title}{About specific features of the structure modulation in
  the {Bi-ferrate} compounds isostructural with
  {Bi$_2$Sr$_2$CaCu$_2$O$_8$}},{''} \bibinfo{journal}{Physica C}
  \textbf{\bibinfo{volume}{390}}, \bibinfo{pages}{311} (\bibinfo{year}{2003}).

\bibitem[{\citenamefont{Lambert et~al.}(2001)\citenamefont{Lambert, Leligny,
  and Grebille}}]{LambertJSSCh2001}
\bibinfo{author}{\bibfnamefont{S.}~\bibnamefont{Lambert}},
  \bibinfo{author}{\bibfnamefont{H.}~\bibnamefont{Leligny}}, \bibnamefont{and}
  \bibinfo{author}{\bibfnamefont{D.}~\bibnamefont{Grebille}},
  {``}\bibinfo{title}{Three forms of the misfit layered cobaltite
  {[Ca$_2$CoO$_3$][CoO$_2$]$_{1.62}$. A 4D} structural investigation},{''}
  \bibinfo{journal}{J. Solid State Chem.} \textbf{\bibinfo{volume}{160}},
  \bibinfo{pages}{322} (\bibinfo{year}{2001}).

\bibitem[{\citenamefont{Grebille et~al.}(2007)\citenamefont{Grebille, Muguerra,
  P{\'{e}}rez, Guilmeau, Rousseli{\`{e}}re, and Funahashi}}]{GrebilleAcCrB2007}
\bibinfo{author}{\bibfnamefont{D.}~\bibnamefont{Grebille}},
  \bibinfo{author}{\bibfnamefont{H.}~\bibnamefont{Muguerra}},
  \bibinfo{author}{\bibfnamefont{O.}~\bibnamefont{P{\'{e}}rez}},
  \bibinfo{author}{\bibfnamefont{E.}~\bibnamefont{Guilmeau}},
  \bibinfo{author}{\bibfnamefont{H.}~\bibnamefont{Rousseli{\`{e}}re}},
  \bibnamefont{and}
  \bibinfo{author}{\bibfnamefont{R.}~\bibnamefont{Funahashi}},
  {``}\bibinfo{title}{Superspace crystal symmetry of thermoelectric misfit
  cobalt oxides and predicted structural models},{''} \bibinfo{journal}{Acta
  Cryst. B} \textbf{\bibinfo{volume}{63}}, \bibinfo{pages}{373}
  (\bibinfo{year}{2007}).

\bibitem[{\citenamefont{Kim et~al.}(2012)\citenamefont{Kim, Kim, Choi, Soo~Lim,
  Seo, and Jin~Hwang}}]{KimJAP2012}
\bibinfo{author}{\bibfnamefont{J.-Y.} \bibnamefont{Kim}},
  \bibinfo{author}{\bibfnamefont{J.-I.} \bibnamefont{Kim}},
  \bibinfo{author}{\bibfnamefont{S.-M.} \bibnamefont{Choi}},
  \bibinfo{author}{\bibfnamefont{Y.}~\bibnamefont{Soo~Lim}},
  \bibinfo{author}{\bibfnamefont{W.-S.} \bibnamefont{Seo}}, \bibnamefont{and}
  \bibinfo{author}{\bibfnamefont{H.}~\bibnamefont{Jin~Hwang}},
  {``}\bibinfo{title}{Nanostructured thermoelectric cobalt oxide by
  exfoliation/restacking route},{''} \bibinfo{journal}{J. Appl. Phys.}
  \textbf{\bibinfo{volume}{112}}, \bibinfo{pages}{113705}
  (\bibinfo{year}{2012}).

\bibitem[{\citenamefont{Gavrichkov et~al.}(2019)\citenamefont{Gavrichkov,
  Shan'ko, Zamkova, and Bianconi}}]{GavrichkovJPhChL2019}
\bibinfo{author}{\bibfnamefont{V.~A.} \bibnamefont{Gavrichkov}},
  \bibinfo{author}{\bibfnamefont{Y.}~\bibnamefont{Shan'ko}},
  \bibinfo{author}{\bibfnamefont{N.~G.} \bibnamefont{Zamkova}},
  \bibnamefont{and} \bibinfo{author}{\bibfnamefont{A.}~\bibnamefont{Bianconi}},
  {``}\bibinfo{title}{Is there any hidden symmetry in the stripe structure of
  perovskite high-temperature superconductors?},{''} \bibinfo{journal}{J. Phys.
  Chem. Lett.} \textbf{\bibinfo{volume}{10}}, \bibinfo{pages}{1840}
  (\bibinfo{year}{2019}).

\bibitem[{\citenamefont{Mazza et~al.}(2022)\citenamefont{Mazza, Gao, Rossi,
  Musico, Valentine, Kennedy, Zhang, Lapano, Keppens, Moore
  et~al.}}]{MazzaJVacSci2022}
\bibinfo{author}{\bibfnamefont{A.~R.} \bibnamefont{Mazza}},
  \bibinfo{author}{\bibfnamefont{X.}~\bibnamefont{Gao}},
  \bibinfo{author}{\bibfnamefont{D.~J.} \bibnamefont{Rossi}},
  \bibinfo{author}{\bibfnamefont{B.~L.} \bibnamefont{Musico}},
  \bibinfo{author}{\bibfnamefont{T.~W.} \bibnamefont{Valentine}},
  \bibinfo{author}{\bibfnamefont{Z.}~\bibnamefont{Kennedy}},
  \bibinfo{author}{\bibfnamefont{J.}~\bibnamefont{Zhang}},
  \bibinfo{author}{\bibfnamefont{J.}~\bibnamefont{Lapano}},
  \bibinfo{author}{\bibfnamefont{V.}~\bibnamefont{Keppens}},
  \bibinfo{author}{\bibfnamefont{R.~G.} \bibnamefont{Moore}},
  \bibnamefont{et~al.}, {``}\bibinfo{title}{Searching for superconductivity in
  high entropy oxide {Ruddlesden--Popper} cuprate films},{''}
  \bibinfo{journal}{J. Vac. Sci. Tech. A} \textbf{\bibinfo{volume}{40}},
  \bibinfo{pages}{013404} (\bibinfo{year}{2022}).

\bibitem[{\citenamefont{Kaneko et~al.}(2004)\citenamefont{Kaneko, Shimizu,
  Akiyama, Ito, Mitsuhashi, Ohya, Saito, Funakubo, and
  Yoshimoto}}]{KanekoAPL2004}
\bibinfo{author}{\bibfnamefont{S.}~\bibnamefont{Kaneko}},
  \bibinfo{author}{\bibfnamefont{Y.}~\bibnamefont{Shimizu}},
  \bibinfo{author}{\bibfnamefont{K.}~\bibnamefont{Akiyama}},
  \bibinfo{author}{\bibfnamefont{T.}~\bibnamefont{Ito}},
  \bibinfo{author}{\bibfnamefont{M.}~\bibnamefont{Mitsuhashi}},
  \bibinfo{author}{\bibfnamefont{S.}~\bibnamefont{Ohya}},
  \bibinfo{author}{\bibfnamefont{K.}~\bibnamefont{Saito}},
  \bibinfo{author}{\bibfnamefont{H.}~\bibnamefont{Funakubo}}, \bibnamefont{and}
  \bibinfo{author}{\bibfnamefont{M.}~\bibnamefont{Yoshimoto}},
  {``}\bibinfo{title}{Modulation derived satellite peaks in x-ray reciprocal
  mapping on bismuth cuprate superconductor film},{''} \bibinfo{journal}{Appl.
  Phys. Lett.} \textbf{\bibinfo{volume}{85}}, \bibinfo{pages}{2301}
  (\bibinfo{year}{2004}).

\bibitem[{\citenamefont{Poccia et~al.}(2011)\citenamefont{Poccia, Campi,
  Fratini, Ricci, Saini, and Bianconi}}]{PocciaPRB2011}
\bibinfo{author}{\bibfnamefont{N.}~\bibnamefont{Poccia}},
  \bibinfo{author}{\bibfnamefont{G.}~\bibnamefont{Campi}},
  \bibinfo{author}{\bibfnamefont{M.}~\bibnamefont{Fratini}},
  \bibinfo{author}{\bibfnamefont{A.}~\bibnamefont{Ricci}},
  \bibinfo{author}{\bibfnamefont{N.~L.} \bibnamefont{Saini}}, \bibnamefont{and}
  \bibinfo{author}{\bibfnamefont{A.}~\bibnamefont{Bianconi}},
  {``}\bibinfo{title}{Spatial inhomogeneity and planar symmetry breaking of the
  lattice incommensurate supermodulation in the high-temperature superconductor
  Bi$_{2}$Sr$_{2}$CaCu$_{2}$O$_{8+y}$},{''} \bibinfo{journal}{Phys. Rev. B}
  \textbf{\bibinfo{volume}{84}}, \bibinfo{pages}{100504}
  (\bibinfo{year}{2011}).

\bibitem[{\citenamefont{Poccia et~al.}(2020)\citenamefont{Poccia, Zhao, Yoo,
  Huang, Yan, Chu, Zhong, Gu, Mazzoli, Watanabe et~al.}}]{PocciaPRM2020}
\bibinfo{author}{\bibfnamefont{N.}~\bibnamefont{Poccia}},
  \bibinfo{author}{\bibfnamefont{S.~Y.~F.} \bibnamefont{Zhao}},
  \bibinfo{author}{\bibfnamefont{H.}~\bibnamefont{Yoo}},
  \bibinfo{author}{\bibfnamefont{X.}~\bibnamefont{Huang}},
  \bibinfo{author}{\bibfnamefont{H.}~\bibnamefont{Yan}},
  \bibinfo{author}{\bibfnamefont{Y.~S.} \bibnamefont{Chu}},
  \bibinfo{author}{\bibfnamefont{R.}~\bibnamefont{Zhong}},
  \bibinfo{author}{\bibfnamefont{G.}~\bibnamefont{Gu}},
  \bibinfo{author}{\bibfnamefont{C.}~\bibnamefont{Mazzoli}},
  \bibinfo{author}{\bibfnamefont{K.}~\bibnamefont{Watanabe}},
  \bibnamefont{et~al.}, {``}\bibinfo{title}{Spatially correlated incommensurate
  lattice modulations in an atomically thin high-temperature
  ${\mathrm{Bi}}_{2.1}{\mathrm{Sr}}_{1.9}\mathrm{Ca}{\mathrm{Cu}}_{2.0}{\mathrm{O}}_{8+y}$
  superconductor},{''} \bibinfo{journal}{Phys. Rev. Materials}
  \textbf{\bibinfo{volume}{4}}, \bibinfo{pages}{114007} (\bibinfo{year}{2020}).

\bibitem[{\citenamefont{Zhang et~al.}(2016)\citenamefont{Zhang, Hu, Hu, Zhao,
  Ding, Sun, Liang, Zhang, He, Liu et~al.}}]{ZhangSciBul2016}
\bibinfo{author}{\bibfnamefont{Y.}~\bibnamefont{Zhang}},
  \bibinfo{author}{\bibfnamefont{C.}~\bibnamefont{Hu}},
  \bibinfo{author}{\bibfnamefont{Y.}~\bibnamefont{Hu}},
  \bibinfo{author}{\bibfnamefont{L.}~\bibnamefont{Zhao}},
  \bibinfo{author}{\bibfnamefont{Y.}~\bibnamefont{Ding}},
  \bibinfo{author}{\bibfnamefont{X.}~\bibnamefont{Sun}},
  \bibinfo{author}{\bibfnamefont{A.}~\bibnamefont{Liang}},
  \bibinfo{author}{\bibfnamefont{Y.}~\bibnamefont{Zhang}},
  \bibinfo{author}{\bibfnamefont{S.}~\bibnamefont{He}},
  \bibinfo{author}{\bibfnamefont{D.}~\bibnamefont{Liu}}, \bibnamefont{et~al.},
  {``}\bibinfo{title}{In situ carrier tuning in high temperature superconductor
  {Bi$_2$Sr$_2$CaCu$_2$O$_{8+\delta}$} by potassium deposition},{''}
  \bibinfo{journal}{Sci. Bull.} \textbf{\bibinfo{volume}{61}},
  \bibinfo{pages}{1037} (\bibinfo{year}{2016}).

\bibitem[{\citenamefont{Bianconi and Poccia}(2012)}]{BianconiPocciaJSCNM2012}
\bibinfo{author}{\bibfnamefont{A.}~\bibnamefont{Bianconi}} \bibnamefont{and}
  \bibinfo{author}{\bibfnamefont{N.}~\bibnamefont{Poccia}},
  {``}\bibinfo{title}{Superstripes and complexity in high-temperature
  superconductors},{''} \bibinfo{journal}{J. Supercond. Novel Magn.}
  \textbf{\bibinfo{volume}{25}}, \bibinfo{pages}{1403} (\bibinfo{year}{2012}).

\bibitem[{\citenamefont{Bianconi et~al.}(1993)\citenamefont{Bianconi, Longa,
  Missori, Pettiti, Pompa, and Soldatov}}]{BianconiJpnJAP1993}
\bibinfo{author}{\bibfnamefont{A.}~\bibnamefont{Bianconi}},
  \bibinfo{author}{\bibfnamefont{S.~D.} \bibnamefont{Longa}},
  \bibinfo{author}{\bibfnamefont{M.}~\bibnamefont{Missori}},
  \bibinfo{author}{\bibfnamefont{I.}~\bibnamefont{Pettiti}},
  \bibinfo{author}{\bibfnamefont{M.}~\bibnamefont{Pompa}}, \bibnamefont{and}
  \bibinfo{author}{\bibfnamefont{A.}~\bibnamefont{Soldatov}},
  {``}\bibinfo{title}{Structure of the different {Cu} sites in the corrugated
  {CuO$_2$} plane in high {$T_c$} superconductors},{''} \bibinfo{journal}{Jpn.
  J. Appl. Phys.} \textbf{\bibinfo{volume}{32}}, \bibinfo{pages}{578}
  (\bibinfo{year}{1993}).

\bibitem[{\citenamefont{Alldredge et~al.}(2008)\citenamefont{Alldredge, Lee,
  McElroy, Wang, Fujita, Kohsaka, Taylor, Eisaki, Uchida, and
  Davis}}]{AlldredgeNatPh2008}
\bibinfo{author}{\bibfnamefont{J.}~\bibnamefont{Alldredge}},
  \bibinfo{author}{\bibfnamefont{J.}~\bibnamefont{Lee}},
  \bibinfo{author}{\bibfnamefont{K.}~\bibnamefont{McElroy}},
  \bibinfo{author}{\bibfnamefont{M.}~\bibnamefont{Wang}},
  \bibinfo{author}{\bibfnamefont{K.}~\bibnamefont{Fujita}},
  \bibinfo{author}{\bibfnamefont{Y.}~\bibnamefont{Kohsaka}},
  \bibinfo{author}{\bibfnamefont{C.}~\bibnamefont{Taylor}},
  \bibinfo{author}{\bibfnamefont{H.}~\bibnamefont{Eisaki}},
  \bibinfo{author}{\bibfnamefont{P.}~\bibnamefont{Uchida},
  \bibfnamefont{S.~Hirschfeld}}, \bibnamefont{and}
  \bibinfo{author}{\bibfnamefont{J.}~\bibnamefont{Davis}},
  {``}\bibinfo{title}{Evolution of the electronic excitation spectrum with
  strongly diminishing hole density in superconducting
  {Bi$_2$Sr$_2$CaCu$_2$O$_{8 + \delta}$}},{''} \bibinfo{journal}{Nat. Phys.}
  \textbf{\bibinfo{volume}{4}}, \bibinfo{pages}{319} (\bibinfo{year}{2008}).

\bibitem[{\citenamefont{Pasupathy et~al.}(2008)\citenamefont{Pasupathy, Pushp,
  Gomes, Parker, Wen, Xu, Gu, Ono, Ando, and Yazdani}}]{PasupathyScience2008}
\bibinfo{author}{\bibfnamefont{A.~N.} \bibnamefont{Pasupathy}},
  \bibinfo{author}{\bibfnamefont{A.}~\bibnamefont{Pushp}},
  \bibinfo{author}{\bibfnamefont{K.~K.} \bibnamefont{Gomes}},
  \bibinfo{author}{\bibfnamefont{C.~V.} \bibnamefont{Parker}},
  \bibinfo{author}{\bibfnamefont{J.}~\bibnamefont{Wen}},
  \bibinfo{author}{\bibfnamefont{Z.}~\bibnamefont{Xu}},
  \bibinfo{author}{\bibfnamefont{G.}~\bibnamefont{Gu}},
  \bibinfo{author}{\bibfnamefont{S.}~\bibnamefont{Ono}},
  \bibinfo{author}{\bibfnamefont{Y.}~\bibnamefont{Ando}}, \bibnamefont{and}
  \bibinfo{author}{\bibfnamefont{A.}~\bibnamefont{Yazdani}},
  {``}\bibinfo{title}{Electronic origin of the inhomogeneous pairing
  interaction in the {high-$T_c$} superconductor
  {Bi$_2$Sr$_2$CaCu$_2$O$_{8+\delta}$}},{''} \bibinfo{journal}{Science}
  \textbf{\bibinfo{volume}{320}}, \bibinfo{pages}{196} (\bibinfo{year}{2008}).

\bibitem[{\citenamefont{Gomes et~al.}(2007)\citenamefont{Gomes, Pasupathy,
  Pushp, Ono, Ando, and Yazdani}}]{GomesNature2007}
\bibinfo{author}{\bibfnamefont{K.~K.} \bibnamefont{Gomes}},
  \bibinfo{author}{\bibfnamefont{A.~N.} \bibnamefont{Pasupathy}},
  \bibinfo{author}{\bibfnamefont{A.}~\bibnamefont{Pushp}},
  \bibinfo{author}{\bibfnamefont{S.}~\bibnamefont{Ono}},
  \bibinfo{author}{\bibfnamefont{Y.}~\bibnamefont{Ando}}, \bibnamefont{and}
  \bibinfo{author}{\bibfnamefont{A.}~\bibnamefont{Yazdani}},
  {``}\bibinfo{title}{Visualizing pair formation on the atomic scale in the
  {high-$T_c$} superconductor {Bi$_2$Sr$_2$CaCu$_2$O$_{8+\delta}$}},{''}
  \bibinfo{journal}{Nature} \textbf{\bibinfo{volume}{447}},
  \bibinfo{pages}{569} (\bibinfo{year}{2007}).

\bibitem[{\citenamefont{Bo\v{z}ovi\'{c}
  et~al.}(2016)\citenamefont{Bo\v{z}ovi\'{c}, He, Wu, and
  Bollinger}}]{BozovicNature2016}
\bibinfo{author}{\bibfnamefont{I.}~\bibnamefont{Bo\v{z}ovi\'{c}}},
  \bibinfo{author}{\bibfnamefont{X.}~\bibnamefont{He}},
  \bibinfo{author}{\bibfnamefont{J.}~\bibnamefont{Wu}}, \bibnamefont{and}
  \bibinfo{author}{\bibfnamefont{A.}~\bibnamefont{Bollinger}},
  {``}\bibinfo{title}{Dependence of the critical temperature in overdoped
  copper oxides on superfluid density},{''} \bibinfo{journal}{Nature}
  \textbf{\bibinfo{volume}{536}}, \bibinfo{pages}{309} (\bibinfo{year}{2016}).

\bibitem[{\citenamefont{He et~al.}(2021)\citenamefont{He, Chen, Li, Zhao, Song,
  Yoshida, Eisaki, Wu, Chen, Lu et~al.}}]{HePRX2021}
\bibinfo{author}{\bibfnamefont{Y.}~\bibnamefont{He}},
  \bibinfo{author}{\bibfnamefont{S.-D.} \bibnamefont{Chen}},
  \bibinfo{author}{\bibfnamefont{Z.-X.} \bibnamefont{Li}},
  \bibinfo{author}{\bibfnamefont{D.}~\bibnamefont{Zhao}},
  \bibinfo{author}{\bibfnamefont{D.}~\bibnamefont{Song}},
  \bibinfo{author}{\bibfnamefont{Y.}~\bibnamefont{Yoshida}},
  \bibinfo{author}{\bibfnamefont{H.}~\bibnamefont{Eisaki}},
  \bibinfo{author}{\bibfnamefont{T.}~\bibnamefont{Wu}},
  \bibinfo{author}{\bibfnamefont{X.-H.} \bibnamefont{Chen}},
  \bibinfo{author}{\bibfnamefont{D.-H.} \bibnamefont{Lu}},
  \bibnamefont{et~al.}, {``}\bibinfo{title}{Superconducting fluctuations in
  overdoped {Bi$_2$Sr$_2$CaCu$_2$O$_{8+{\delta}}$}},{''}
  \bibinfo{journal}{Phys. Rev. X} \textbf{\bibinfo{volume}{11}},
  \bibinfo{pages}{031068} (\bibinfo{year}{2021}).

\bibitem[{\citenamefont{Drozdov et~al.}(2018)\citenamefont{Drozdov,
  Pletikosi\'{c}, Kim, Fujita, Gu, Davis, Johnson, Bo\v{z}ovi\'{c}, and
  Valla}}]{DrozdovNature2018}
\bibinfo{author}{\bibfnamefont{I.~K.} \bibnamefont{Drozdov}},
  \bibinfo{author}{\bibfnamefont{I.}~\bibnamefont{Pletikosi\'{c}}},
  \bibinfo{author}{\bibfnamefont{C.-K.} \bibnamefont{Kim}},
  \bibinfo{author}{\bibfnamefont{K.}~\bibnamefont{Fujita}},
  \bibinfo{author}{\bibfnamefont{G.}~\bibnamefont{Gu}},
  \bibinfo{author}{\bibfnamefont{J.~C.~S.} \bibnamefont{Davis}},
  \bibinfo{author}{\bibfnamefont{P.}~\bibnamefont{Johnson}},
  \bibinfo{author}{\bibfnamefont{I.}~\bibnamefont{Bo\v{z}ovi\'{c}}},
  \bibnamefont{and} \bibinfo{author}{\bibfnamefont{T.}~\bibnamefont{Valla}},
  {``}\bibinfo{title}{Phase diagram of {Bi$_2$Sr$_2$CaCu$_2$O$_{8+\delta}$}
  revisited},{''} \bibinfo{journal}{Nat. Commun.} \textbf{\bibinfo{volume}{9}},
  \bibinfo{pages}{5210} (\bibinfo{year}{2018}).

\bibitem[{\citenamefont{Li et~al.}(2018)\citenamefont{Li, Ding, He, Ruan, Cai,
  Ye, Hao, Zhao, Zhou, Wang et~al.}}]{LiNJP2018}
\bibinfo{author}{\bibfnamefont{X.}~\bibnamefont{Li}},
  \bibinfo{author}{\bibfnamefont{Y.}~\bibnamefont{Ding}},
  \bibinfo{author}{\bibfnamefont{C.}~\bibnamefont{He}},
  \bibinfo{author}{\bibfnamefont{W.}~\bibnamefont{Ruan}},
  \bibinfo{author}{\bibfnamefont{P.}~\bibnamefont{Cai}},
  \bibinfo{author}{\bibfnamefont{C.}~\bibnamefont{Ye}},
  \bibinfo{author}{\bibfnamefont{Z.}~\bibnamefont{Hao}},
  \bibinfo{author}{\bibfnamefont{L.}~\bibnamefont{Zhao}},
  \bibinfo{author}{\bibfnamefont{X.}~\bibnamefont{Zhou}},
  \bibinfo{author}{\bibfnamefont{Q.}~\bibnamefont{Wang}}, \bibnamefont{et~al.},
  {``}\bibinfo{title}{Quasiparticle interference and charge order in a heavily
  overdoped non-superconducting cuprate},{''} \bibinfo{journal}{New J. Phys.}
  \textbf{\bibinfo{volume}{20}}, \bibinfo{pages}{063041}
  (\bibinfo{year}{2018}).

\bibitem[{\citenamefont{Maier et~al.}(2020)\citenamefont{Maier, Karakuzu, and
  Scalapino}}]{MaierPRRes2020}
\bibinfo{author}{\bibfnamefont{T.~A.} \bibnamefont{Maier}},
  \bibinfo{author}{\bibfnamefont{S.}~\bibnamefont{Karakuzu}}, \bibnamefont{and}
  \bibinfo{author}{\bibfnamefont{D.~J.} \bibnamefont{Scalapino}},
  {``}\bibinfo{title}{Overdoped end of the cuprate phase diagram},{''}
  \bibinfo{journal}{Phys. Rev. Research} \textbf{\bibinfo{volume}{2}},
  \bibinfo{pages}{033132} (\bibinfo{year}{2020}).

\bibitem[{\citenamefont{{M.Yu. Kagan} et~al.}(2021)\citenamefont{{M.Yu. Kagan},
  {K.I. Kugel}, and {A.L. Rakhmanov}}}]{KaganPhRep2021}
\bibinfo{author}{\bibnamefont{{M.Yu. Kagan}}},
  \bibinfo{author}{\bibnamefont{{K.I. Kugel}}}, \bibnamefont{and}
  \bibinfo{author}{\bibnamefont{{A.L. Rakhmanov}}},
  {``}\bibinfo{title}{Electronic phase separation: {Recent} progress in the old
  problem},{''} \bibinfo{journal}{Phys. Rep.} \textbf{\bibinfo{volume}{916}},
  \bibinfo{pages}{1} (\bibinfo{year}{2021}).

\bibitem[{\citenamefont{Rakhmanov et~al.}(2017)\citenamefont{Rakhmanov, Kugel,
  {M. Yu. Kagan}, Rozhkov, and Sboychakov}}]{RakhmanovJETPL2017}
\bibinfo{author}{\bibfnamefont{A.~L.} \bibnamefont{Rakhmanov}},
  \bibinfo{author}{\bibfnamefont{K.~I.} \bibnamefont{Kugel}},
  \bibinfo{author}{\bibnamefont{{M. Yu. Kagan}}},
  \bibinfo{author}{\bibfnamefont{A.~V.} \bibnamefont{Rozhkov}},
  \bibnamefont{and} \bibinfo{author}{\bibfnamefont{A.~O.}
  \bibnamefont{Sboychakov}}, {``}\bibinfo{title}{Inhomogeneous electron states
  in the systems with imperfect nesting},{''} \bibinfo{journal}{JETP Letters}
  \textbf{\bibinfo{volume}{105}}, \bibinfo{pages}{806} (\bibinfo{year}{2017}).

\bibitem[{\citenamefont{Kugel et~al.}(2008{\natexlab{a}})\citenamefont{Kugel,
  Rakhmanov, Sboychakov, Poccia, and Bianconi}}]{KugelPRB2008}
\bibinfo{author}{\bibfnamefont{K.~I.} \bibnamefont{Kugel}},
  \bibinfo{author}{\bibfnamefont{A.~L.} \bibnamefont{Rakhmanov}},
  \bibinfo{author}{\bibfnamefont{A.~O.} \bibnamefont{Sboychakov}},
  \bibinfo{author}{\bibfnamefont{N.}~\bibnamefont{Poccia}}, \bibnamefont{and}
  \bibinfo{author}{\bibfnamefont{A.}~\bibnamefont{Bianconi}},
  {``}\bibinfo{title}{Model for phase separation controlled by doping and the
  internal chemical pressure in different cuprate superconductors},{''}
  \bibinfo{journal}{Phys. Rev. B} \textbf{\bibinfo{volume}{78}},
  \bibinfo{pages}{165124} (\bibinfo{year}{2008}{\natexlab{a}}).

\bibitem[{\citenamefont{Bianconi et~al.}(2015)\citenamefont{Bianconi, Poccia,
  Sboychakov, Rakhmanov, and Kugel}}]{BianconiSST2015}
\bibinfo{author}{\bibfnamefont{A.}~\bibnamefont{Bianconi}},
  \bibinfo{author}{\bibfnamefont{N.}~\bibnamefont{Poccia}},
  \bibinfo{author}{\bibfnamefont{A.~O.} \bibnamefont{Sboychakov}},
  \bibinfo{author}{\bibfnamefont{A.~L.} \bibnamefont{Rakhmanov}},
  \bibnamefont{and} \bibinfo{author}{\bibfnamefont{K.~I.} \bibnamefont{Kugel}},
  {``}\bibinfo{title}{Intrinsic arrested nanoscale phase separation near a
  topological {Lifshitz} transition in strongly correlated two-band
  metals},{''} \bibinfo{journal}{Supercond. Sci. Technol.}
  \textbf{\bibinfo{volume}{28}}, \bibinfo{pages}{024005}
  (\bibinfo{year}{2015}).

\bibitem[{\citenamefont{Kugel et~al.}(2008{\natexlab{b}})\citenamefont{Kugel,
  Rakhmanov, Sboychakov, Kusmartsev, Poccia, and Bianconi}}]{KugelSST2008}
\bibinfo{author}{\bibfnamefont{K.~I.} \bibnamefont{Kugel}},
  \bibinfo{author}{\bibfnamefont{A.~L.} \bibnamefont{Rakhmanov}},
  \bibinfo{author}{\bibfnamefont{A.~O.} \bibnamefont{Sboychakov}},
  \bibinfo{author}{\bibfnamefont{F.~V.} \bibnamefont{Kusmartsev}},
  \bibinfo{author}{\bibfnamefont{N.}~\bibnamefont{Poccia}}, \bibnamefont{and}
  \bibinfo{author}{\bibfnamefont{A.}~\bibnamefont{Bianconi}},
  {``}\bibinfo{title}{A two-band model for the phase separation induced by the
  chemical mismatch pressure in different cuprate superconductors},{''}
  \bibinfo{journal}{Supercond. Sci. Technol.} \textbf{\bibinfo{volume}{22}},
  \bibinfo{pages}{014007} (\bibinfo{year}{2008}{\natexlab{b}}).

\bibitem[{\citenamefont{Massidda et~al.}(1988)\citenamefont{Massidda, Yu, and
  Freeman}}]{MassiddaPhC1988}
\bibinfo{author}{\bibfnamefont{S.}~\bibnamefont{Massidda}},
  \bibinfo{author}{\bibfnamefont{J.}~\bibnamefont{Yu}}, \bibnamefont{and}
  \bibinfo{author}{\bibfnamefont{A.}~\bibnamefont{Freeman}},
  {``}\bibinfo{title}{Electronic structure and properties of
  {Bi$_2$Sr$_2$CaCu$_2$O$_8$}, the third {high-$T_c$} superconductor},{''}
  \bibinfo{journal}{Physica C} \textbf{\bibinfo{volume}{152}},
  \bibinfo{pages}{251} (\bibinfo{year}{1988}).

\bibitem[{\citenamefont{Nokelainen et~al.}(2020)\citenamefont{Nokelainen, Lane,
  Markiewicz, Barbiellini, Pulkkinen, Singh, Sun, Pussi, and
  Bansil}}]{NokelainenPRB2020}
\bibinfo{author}{\bibfnamefont{J.}~\bibnamefont{Nokelainen}},
  \bibinfo{author}{\bibfnamefont{C.}~\bibnamefont{Lane}},
  \bibinfo{author}{\bibfnamefont{R.~S.} \bibnamefont{Markiewicz}},
  \bibinfo{author}{\bibfnamefont{B.}~\bibnamefont{Barbiellini}},
  \bibinfo{author}{\bibfnamefont{A.}~\bibnamefont{Pulkkinen}},
  \bibinfo{author}{\bibfnamefont{B.}~\bibnamefont{Singh}},
  \bibinfo{author}{\bibfnamefont{J.}~\bibnamefont{Sun}},
  \bibinfo{author}{\bibfnamefont{K.}~\bibnamefont{Pussi}}, \bibnamefont{and}
  \bibinfo{author}{\bibfnamefont{A.}~\bibnamefont{Bansil}},
  {``}\bibinfo{title}{Ab initio description of the
  {Bi$_2$Sr$_2$CaCu$_2$O$_{8+\delta}$} electronic structure},{''}
  \bibinfo{journal}{Phys. Rev. B} \textbf{\bibinfo{volume}{101}},
  \bibinfo{pages}{214523} (\bibinfo{year}{2020}).

\bibitem[{\citenamefont{Hewat et~al.}(1989)\citenamefont{Hewat, Capponi, and
  Marezio}}]{HewatPhC1989}
\bibinfo{author}{\bibfnamefont{E.}~\bibnamefont{Hewat}},
  \bibinfo{author}{\bibfnamefont{J.}~\bibnamefont{Capponi}}, \bibnamefont{and}
  \bibinfo{author}{\bibfnamefont{M.}~\bibnamefont{Marezio}},
  {``}\bibinfo{title}{A model for the superstructure of
  {Bi$_2$Sr$_2$CaCu$_2$O$_{8.2}$}},{''} \bibinfo{journal}{Physica C}
  \textbf{\bibinfo{volume}{157}}, \bibinfo{pages}{502} (\bibinfo{year}{1989}).

\bibitem[{\citenamefont{Bak}(1982)}]{BakRPPh1982}
\bibinfo{author}{\bibfnamefont{P.}~\bibnamefont{Bak}},
  {``}\bibinfo{title}{Commensurate phases, incommensurate phases and the
  devil's staircase},{''} \bibinfo{journal}{Rep. Prog. Phys.}
  \textbf{\bibinfo{volume}{45}}, \bibinfo{pages}{587} (\bibinfo{year}{1982}).

\bibitem[{\citenamefont{Andersen et~al.}(1995)\citenamefont{Andersen,
  Liechtenstein, Jepsen, and Paulsen}}]{LDAenergyBands1995}
\bibinfo{author}{\bibfnamefont{O.}~\bibnamefont{Andersen}},
  \bibinfo{author}{\bibfnamefont{A.}~\bibnamefont{Liechtenstein}},
  \bibinfo{author}{\bibfnamefont{O.}~\bibnamefont{Jepsen}}, \bibnamefont{and}
  \bibinfo{author}{\bibfnamefont{F.}~\bibnamefont{Paulsen}},
  {``}\bibinfo{title}{{LDA} energy bands, low-energy {Hamiltonians}, $t'$,
  $t''$ $t_{\bot}(k)$, and $J_{\bot}$},{''} \bibinfo{journal}{J. Phys. Chem.
  Solids} \textbf{\bibinfo{volume}{56}}, \bibinfo{pages}{1573}
  (\bibinfo{year}{1995}).

\bibitem[{\citenamefont{Slater and Koster}(1954)}]{SlaterKoster}
\bibinfo{author}{\bibfnamefont{J.~C.} \bibnamefont{Slater}} \bibnamefont{and}
  \bibinfo{author}{\bibfnamefont{G.~F.} \bibnamefont{Koster}},
  {``}\bibinfo{title}{Simplified {LCAO} method for the periodic potential
  problem},{''} \bibinfo{journal}{Phys. Rev.} \textbf{\bibinfo{volume}{94}},
  \bibinfo{pages}{1498} (\bibinfo{year}{1954}).

\bibitem[{\citenamefont{Bianconi
  et~al.}(1996{\natexlab{d}})\citenamefont{Bianconi, Lusignoli, Saini, Bordet,
  Kvick, and Radaelli}}]{BianLusPRB1996}
\bibinfo{author}{\bibfnamefont{A.}~\bibnamefont{Bianconi}},
  \bibinfo{author}{\bibfnamefont{M.}~\bibnamefont{Lusignoli}},
  \bibinfo{author}{\bibfnamefont{N.~L.} \bibnamefont{Saini}},
  \bibinfo{author}{\bibfnamefont{P.}~\bibnamefont{Bordet}},
  \bibinfo{author}{\bibfnamefont{A.}~\bibnamefont{Kvick}}, \bibnamefont{and}
  \bibinfo{author}{\bibfnamefont{P.~G.} \bibnamefont{Radaelli}},
  {``}\bibinfo{title}{Stripe structure of the CuO$_2$ plane in
  Bi$_2$Sr$_2$CaCu$_2$O$_{8 + y}$ by anomalous x-ray diffraction},{''}
  \bibinfo{journal}{Phys. Rev. B} \textbf{\bibinfo{volume}{54}},
  \bibinfo{pages}{4310} (\bibinfo{year}{1996}{\natexlab{d}}).

\bibitem[{\citenamefont{Pavarini et~al.}(2001)\citenamefont{Pavarini, Dasgupta,
  Saha-Dasgupta, Jepsen, and Andersen}}]{PavariniPRL2001}
\bibinfo{author}{\bibfnamefont{E.}~\bibnamefont{Pavarini}},
  \bibinfo{author}{\bibfnamefont{I.}~\bibnamefont{Dasgupta}},
  \bibinfo{author}{\bibfnamefont{T.}~\bibnamefont{Saha-Dasgupta}},
  \bibinfo{author}{\bibfnamefont{O.}~\bibnamefont{Jepsen}}, \bibnamefont{and}
  \bibinfo{author}{\bibfnamefont{O.~K.} \bibnamefont{Andersen}},
  {``}\bibinfo{title}{Band-Structure Trend in Hole-Doped Cuprates and
  Correlation with ${\mathit{T}}_{\mathit{c}\mathrm{max}}$},{''}
  \bibinfo{journal}{Phys. Rev. Lett.} \textbf{\bibinfo{volume}{87}},
  \bibinfo{pages}{047003} (\bibinfo{year}{2001}).

\bibitem[{\citenamefont{Yuan et~al.}(2019)\citenamefont{Yuan, Isobe, and
  Fu}}]{YuanNatCom2019}
\bibinfo{author}{\bibfnamefont{N.~F.~Q.} \bibnamefont{Yuan}},
  \bibinfo{author}{\bibfnamefont{H.}~\bibnamefont{Isobe}}, \bibnamefont{and}
  \bibinfo{author}{\bibfnamefont{L.}~\bibnamefont{Fu}},
  {``}\bibinfo{title}{Magic of high-order van {Hove} singularity},{''}
  \bibinfo{journal}{Nat. Commun.} \textbf{\bibinfo{volume}{10}},
  \bibinfo{pages}{5769} (\bibinfo{year}{2019}).

\bibitem[{\citenamefont{Isobe and Fu}(2019)}]{IsobePRR2019}
\bibinfo{author}{\bibfnamefont{H.}~\bibnamefont{Isobe}} \bibnamefont{and}
  \bibinfo{author}{\bibfnamefont{L.}~\bibnamefont{Fu}},
  {``}\bibinfo{title}{Supermetal},{''} \bibinfo{journal}{Phys. Rev. Research}
  \textbf{\bibinfo{volume}{1}}, \bibinfo{pages}{033206} (\bibinfo{year}{2019}).

\bibitem[{\citenamefont{Classen et~al.}(2020)\citenamefont{Classen, Chubukov,
  Honerkamp, and Scherer}}]{ClassenPRB2019}
\bibinfo{author}{\bibfnamefont{L.}~\bibnamefont{Classen}},
  \bibinfo{author}{\bibfnamefont{A.~V.} \bibnamefont{Chubukov}},
  \bibinfo{author}{\bibfnamefont{C.}~\bibnamefont{Honerkamp}},
  \bibnamefont{and} \bibinfo{author}{\bibfnamefont{M.~M.}
  \bibnamefont{Scherer}}, {``}\bibinfo{title}{Competing orders at higher-order
  {Van Hove} points},{''} \bibinfo{journal}{Phys. Rev. B}
  \textbf{\bibinfo{volume}{102}}, \bibinfo{pages}{125141}
  (\bibinfo{year}{2020}).

\bibitem[{\citenamefont{Guerci et~al.}(2022)\citenamefont{Guerci, Simon, and
  Mora}}]{GuerciPRR2022}
\bibinfo{author}{\bibfnamefont{D.}~\bibnamefont{Guerci}},
  \bibinfo{author}{\bibfnamefont{P.}~\bibnamefont{Simon}}, \bibnamefont{and}
  \bibinfo{author}{\bibfnamefont{C.}~\bibnamefont{Mora}},
  {``}\bibinfo{title}{Higher-order {Van Hove} singularity in magic-angle
  twisted trilayer graphene},{''} \bibinfo{journal}{Phys. Rev. Research}
  \textbf{\bibinfo{volume}{4}}, \bibinfo{pages}{L012013}
  (\bibinfo{year}{2022}).

\bibitem[{\citenamefont{Lopes~dos Santos et~al.}(2012)\citenamefont{Lopes~dos
  Santos, Peres, and Castro~Neto}}]{dSPRB}
\bibinfo{author}{\bibfnamefont{J.~M.~B.} \bibnamefont{Lopes~dos Santos}},
  \bibinfo{author}{\bibfnamefont{N.~M.~R.} \bibnamefont{Peres}},
  \bibnamefont{and} \bibinfo{author}{\bibfnamefont{A.~H.}
  \bibnamefont{Castro~Neto}}, {``}\bibinfo{title}{Continuum model of the
  twisted graphene bilayer},{''} \bibinfo{journal}{Phys. Rev. B}
  \textbf{\bibinfo{volume}{86}}, \bibinfo{pages}{155449}
  (\bibinfo{year}{2012}).

\bibitem[{\citenamefont{Sboychakov et~al.}(2015)\citenamefont{Sboychakov,
  Rakhmanov, Rozhkov, and Nori}}]{ourTBLGPRB2015}
\bibinfo{author}{\bibfnamefont{A.~O.} \bibnamefont{Sboychakov}},
  \bibinfo{author}{\bibfnamefont{A.~L.} \bibnamefont{Rakhmanov}},
  \bibinfo{author}{\bibfnamefont{A.~V.} \bibnamefont{Rozhkov}},
  \bibnamefont{and} \bibinfo{author}{\bibfnamefont{F.}~\bibnamefont{Nori}},
  {``}\bibinfo{title}{Electronic spectrum of twisted bilayer graphene},{''}
  \bibinfo{journal}{Phys. Rev. B} \textbf{\bibinfo{volume}{92}},
  \bibinfo{pages}{075402} (\bibinfo{year}{2015}).

\bibitem[{\citenamefont{Mazziotti
  et~al.}(2021{\natexlab{a}})\citenamefont{Mazziotti, Valletta, Raimondi, and
  Bianconi}}]{MazziottiPRB2021}
\bibinfo{author}{\bibfnamefont{M.~V.} \bibnamefont{Mazziotti}},
  \bibinfo{author}{\bibfnamefont{A.}~\bibnamefont{Valletta}},
  \bibinfo{author}{\bibfnamefont{R.}~\bibnamefont{Raimondi}}, \bibnamefont{and}
  \bibinfo{author}{\bibfnamefont{A.}~\bibnamefont{Bianconi}},
  {``}\bibinfo{title}{Multigap superconductivity at an unconventional{
  Lifshitz} transition in a three-dimensional {Rashba} heterostructure at the
  atomic limit},{''} \bibinfo{journal}{Phys. Rev. B}
  \textbf{\bibinfo{volume}{103}}, \bibinfo{pages}{024523}
  (\bibinfo{year}{2021}{\natexlab{a}}).

\bibitem[{\citenamefont{Perali et~al.}(1996)\citenamefont{Perali, Bianconi,
  Lanzara, and Saini}}]{perali1996gap}
\bibinfo{author}{\bibfnamefont{A.}~\bibnamefont{Perali}},
  \bibinfo{author}{\bibfnamefont{A.}~\bibnamefont{Bianconi}},
  \bibinfo{author}{\bibfnamefont{A.}~\bibnamefont{Lanzara}}, \bibnamefont{and}
  \bibinfo{author}{\bibfnamefont{N.~L.} \bibnamefont{Saini}},
  {``}\bibinfo{title}{The gap amplification at a shape resonance in a
  superlattice of quantum stripes: {A mechanism for high Tc}},{''}
  \bibinfo{journal}{Solid State Commun.} \textbf{\bibinfo{volume}{100}},
  \bibinfo{pages}{181} (\bibinfo{year}{1996}).

\bibitem[{\citenamefont{Perali et~al.}(2012)\citenamefont{Perali, Innocenti,
  Valletta, and Bianconi}}]{PeraliSuST2012}
\bibinfo{author}{\bibfnamefont{A.}~\bibnamefont{Perali}},
  \bibinfo{author}{\bibfnamefont{D.}~\bibnamefont{Innocenti}},
  \bibinfo{author}{\bibfnamefont{A.}~\bibnamefont{Valletta}}, \bibnamefont{and}
  \bibinfo{author}{\bibfnamefont{A.}~\bibnamefont{Bianconi}},
  {``}\bibinfo{title}{Anomalous isotope effect near a 2.5 Lifshitz transition
  in a multi-band multi-condensate superconductor made of a superlattice of
  stripes},{''} \bibinfo{journal}{Supercond. Sci. Technol.}
  \textbf{\bibinfo{volume}{25}}, \bibinfo{pages}{124002}
  (\bibinfo{year}{2012}).

\bibitem[{\citenamefont{Chen et~al.}(2012)\citenamefont{Chen, Shanenko, Perali,
  and Peeters}}]{chen2012superconducting}
\bibinfo{author}{\bibfnamefont{Y.}~\bibnamefont{Chen}},
  \bibinfo{author}{\bibfnamefont{A.~A.} \bibnamefont{Shanenko}},
  \bibinfo{author}{\bibfnamefont{A.}~\bibnamefont{Perali}}, \bibnamefont{and}
  \bibinfo{author}{\bibfnamefont{F.~M.} \bibnamefont{Peeters}},
  {``}\bibinfo{title}{Superconducting nanofilms: molecule-like pairing induced
  by quantum confinement},{''} \bibinfo{journal}{J. Phys.: Condens. Matter}
  \textbf{\bibinfo{volume}{24}}, \bibinfo{pages}{185701}
  (\bibinfo{year}{2012}).

\bibitem[{\citenamefont{Salasnich et~al.}(2019)\citenamefont{Salasnich,
  Shanenko, Vagov, Aguiar, and Perali}}]{salasnich2019screening}
\bibinfo{author}{\bibfnamefont{L.}~\bibnamefont{Salasnich}},
  \bibinfo{author}{\bibfnamefont{A.~A.} \bibnamefont{Shanenko}},
  \bibinfo{author}{\bibfnamefont{A.~A.} \bibnamefont{Vagov}},
  \bibinfo{author}{\bibfnamefont{J.~A.} \bibnamefont{Aguiar}},
  \bibnamefont{and} \bibinfo{author}{\bibfnamefont{A.}~\bibnamefont{Perali}},
  {``}\bibinfo{title}{Screening of pair fluctuations in superconductors with
  coupled shallow and deep {bands: A} route to higher-temperature
  superconductivity},{''} \bibinfo{journal}{Phys. Rev. B}
  \textbf{\bibinfo{volume}{100}}, \bibinfo{pages}{064510}
  (\bibinfo{year}{2019}).

\bibitem[{\citenamefont{Agrestini et~al.}(2004)\citenamefont{Agrestini,
  Metallo, Filippi, Simonelli, Campi, Sanipoli, Liarokapis, De~Negri,
  Giovannini, Saccone et~al.}}]{agrestini2004substitution}
\bibinfo{author}{\bibfnamefont{S.}~\bibnamefont{Agrestini}},
  \bibinfo{author}{\bibfnamefont{C.}~\bibnamefont{Metallo}},
  \bibinfo{author}{\bibfnamefont{M.}~\bibnamefont{Filippi}},
  \bibinfo{author}{\bibfnamefont{L.}~\bibnamefont{Simonelli}},
  \bibinfo{author}{\bibfnamefont{G.}~\bibnamefont{Campi}},
  \bibinfo{author}{\bibfnamefont{C.}~\bibnamefont{Sanipoli}},
  \bibinfo{author}{\bibfnamefont{E.}~\bibnamefont{Liarokapis}},
  \bibinfo{author}{\bibfnamefont{S.}~\bibnamefont{De~Negri}},
  \bibinfo{author}{\bibfnamefont{M.}~\bibnamefont{Giovannini}},
  \bibinfo{author}{\bibfnamefont{A.}~\bibnamefont{Saccone}},
  \bibnamefont{et~al.}, {``}\bibinfo{title}{Substitution {of Sc for Mg in
  MgB$_2$: Effects} on transition temperature and {Kohn} anomaly},{''}
  \bibinfo{journal}{Phys. Rev. B} \textbf{\bibinfo{volume}{70}},
  \bibinfo{pages}{134514} (\bibinfo{year}{2004}).

\bibitem[{\citenamefont{Bianconi et~al.}(2014)\citenamefont{Bianconi,
  Innocenti, Valletta, and Perali}}]{bianconi2014shape}
\bibinfo{author}{\bibfnamefont{A.}~\bibnamefont{Bianconi}},
  \bibinfo{author}{\bibfnamefont{D.}~\bibnamefont{Innocenti}},
  \bibinfo{author}{\bibfnamefont{A.}~\bibnamefont{Valletta}}, \bibnamefont{and}
  \bibinfo{author}{\bibfnamefont{A.}~\bibnamefont{Perali}},
  {``}\bibinfo{title}{Shape resonances in superconducting gaps in a {2DEG} at
  oxide-oxide interface},{''} \bibinfo{journal}{J. Phys. Conf. Ser.}
  \textbf{\bibinfo{volume}{529}}, \bibinfo{pages}{012007}
  (\bibinfo{year}{2014}).

\bibitem[{\citenamefont{Mazziotti
  et~al.}(2021{\natexlab{b}})\citenamefont{Mazziotti, Raimondi, Valletta,
  Campi, and Bianconi}}]{mazziotti2021resonant}
\bibinfo{author}{\bibfnamefont{M.~V.} \bibnamefont{Mazziotti}},
  \bibinfo{author}{\bibfnamefont{R.}~\bibnamefont{Raimondi}},
  \bibinfo{author}{\bibfnamefont{A.}~\bibnamefont{Valletta}},
  \bibinfo{author}{\bibfnamefont{G.}~\bibnamefont{Campi}}, \bibnamefont{and}
  \bibinfo{author}{\bibfnamefont{A.}~\bibnamefont{Bianconi}},
  {``}\bibinfo{title}{Resonant multi-gap superconductivity at room temperature
  near a {Lifshitz} topological transition in sulfur hydrides},{''}
  \bibinfo{journal}{J. Appl. Phys.} \textbf{\bibinfo{volume}{130}},
  \bibinfo{pages}{173904} (\bibinfo{year}{2021}{\natexlab{b}}).

\bibitem[{\citenamefont{Mazziotti
  et~al.}(2021{\natexlab{c}})\citenamefont{Mazziotti, Jarlborg, Bianconi, and
  Valletta}}]{MazziottiEPL2021}
\bibinfo{author}{\bibfnamefont{M.~V.} \bibnamefont{Mazziotti}},
  \bibinfo{author}{\bibfnamefont{T.}~\bibnamefont{Jarlborg}},
  \bibinfo{author}{\bibfnamefont{A.}~\bibnamefont{Bianconi}}, \bibnamefont{and}
  \bibinfo{author}{\bibfnamefont{A.}~\bibnamefont{Valletta}},
  {``}\bibinfo{title}{Room temperature superconductivity dome at a {Fano}
  resonance in superlattices of wires},{''} \bibinfo{journal}{{EPL(Europhysics
  Letters)}} \textbf{\bibinfo{volume}{134}}, \bibinfo{pages}{17001}
  (\bibinfo{year}{2021}{\natexlab{c}}).

\bibitem[{\citenamefont{Tromp et~al.}(2022)\citenamefont{Tromp, Benschop, Ge,
  Battist, Bastiaans, Chatzopoulos, Vervloet, Smit, van Heumen, Golden
  et~al.}}]{tromp2022puddle}
\bibinfo{author}{\bibfnamefont{W.~O.} \bibnamefont{Tromp}},
  \bibinfo{author}{\bibfnamefont{T.}~\bibnamefont{Benschop}},
  \bibinfo{author}{\bibfnamefont{J.-F.} \bibnamefont{Ge}},
  \bibinfo{author}{\bibfnamefont{I.}~\bibnamefont{Battist}},
  \bibinfo{author}{\bibfnamefont{K.~M.} \bibnamefont{Bastiaans}},
  \bibinfo{author}{\bibfnamefont{D.}~\bibnamefont{Chatzopoulos}},
  \bibinfo{author}{\bibfnamefont{A.}~\bibnamefont{Vervloet}},
  \bibinfo{author}{\bibfnamefont{S.}~\bibnamefont{Smit}},
  \bibinfo{author}{\bibfnamefont{E.}~\bibnamefont{van Heumen}},
  \bibinfo{author}{\bibfnamefont{M.~S.} \bibnamefont{Golden}},
  \bibnamefont{et~al.}, {``}\bibinfo{title}{Puddle formation, persistent gaps,
  and non-mean-field breakdown of superconductivity in overdoped
  {(Pb,Bi)$_2$Sr$_2$CuO$_{6+\delta}$}},{''} \bibinfo{journal}{arXiv:2205.09740}
   (\bibinfo{year}{2022}), \eprint{https://arxiv.org/abs/2205.09740}.

\end{thebibliography}

\end{document}